\documentclass[useAMS,usenatbib]{mn2e}
\usepackage{amsmath}
\usepackage{amssymb}
\usepackage{graphicx}
\usepackage{subfigure}
\usepackage{rotating}
\usepackage{fixltx2e}
\usepackage{dblfloatfix}

\title[SED Analysis Through Markov Chains]{SATMC: Spectral Energy
  Distribution Analysis Through Markov Chains}
\author[S.P. Johnson et al.]{
\parbox[t]{\textwidth}{\vspace{-1cm}S.P.~Johnson$^1$, G.W.~Wilson$^1$,
Y.~Tang$^1$, K.S.~Scott$^2$}\\
$^1$Department of Astronomy, University of Massachusetts, Amherst, MA
  01003, USA\\
$^2$North American ALMA Science Center, National Radio Astronomy
Observatory, Charlottesville, VA 22903, USA\\}

\begin{document}
\maketitle

\begin{abstract}
We present the general purpose spectral energy distribution (SED)
fitting tool SED Analysis Through Markov Chains (SATMC).  Utilizing
Monte Carlo Markov Chain (MCMC) algorithms, SATMC fits an observed SED 
to SED templates or models of the user's choice to infer intrinsic 
parameters, generate confidence levels and produce the posterior
parameter distribution.  Here we describe the key features of SATMC 
from the underlying MCMC engine to specific features for handling SED 
fitting.  We detail several test cases of SATMC, comparing results 
obtained to traditional least-squares methods, 
which highlight its accuracy, robustness and wide range of possible 
applications.  We also present a sample of submillimetre galaxies that 
have been fitted using the SED synthesis routine GRASIL as input.  
In general, these SMGs are shown to occupy a large volume of 
parameter space, particularly in regards to their star formation rates 
which range from $\sim$30-3000 M$_{\odot}$ yr$^{-1}$ and stellar masses 
which range from $\sim 10^{10}-10^{12}$ M$_{\odot}$.  Taking advantage of 
the Bayesian formalism inherent to SATMC, we also show how the fitting
results may change under different parametrizations (i.e., different 
initial mass functions) and through additional or improved photometry,
the latter being crucial to the study of high-redshift galaxies.
\end{abstract}

\section{Introduction}

The complete pan-chromatic spectral energy distribution (SED) of a
source encodes a wealth of information concerning its age, mass,
metallicity, dust/gas content, star formation rate (SFR), star
formation history (SFH) and more.  Various emission mechanisms account
for the apparent shape and typical features found in SEDs.  For
example, dust reprocessing results in attenuation of
optical/ultra-violet (UV) photons produced from stellar populations
whose energy is then re-emitted at infrared (IR) wavelengths while
specific species of ions and molecules produce emission/absorption
features across the electromagnetic spectrum. Unfortunately, our
ability to extract information from observations is hampered both on
the observational and theoretical sides.  As observers, we try to make
due with either coarse, broad-band sampling of a source's SED or with
high resolution sampling of a small portion of the SED through
spectroscopy.  Theorists, on the other hand, must make difficult
decisions regarding which physics, spatial scales, and evolutionary
histories to include in the creation of their SED libraries or
synthesis models. 

Within the literature, SED models can be sub-categorized into two main
types.  Empirical models are derived for particular classifications 
based on a subset of similar sources (e.g. Arp 220 and M82 for 
starburst galaxies, the quasar mean template, etc.). These models offer 
the simplest approximation of a source's SED and are generally 
preferred when only sparse photometry is available or for coarse 
estimates of basic properties (e.g. luminosity, SFR, colors). The 
underlying assumption behind empirical models, namely that all sources 
of that 'type' have the same SED, is difficult to verify and so their 
use to probe all but the grossest properties is limited. 

Theoretical models are constructed from sets of physical and radiative 
processes believed to be the dominant contributors to the emission of 
a source.  Functionally, these again are divided into two classes.
Pre-computed  template libraries for particular source types are the
most widely  used \citep[e.g.][and references
  therein]{calzetti00,efstathiou00,siebenmorgen04,sieben07,gawiser09,michalowski10}.
Since they are pre-computed, these libraries allow the rapid 
exploration of a pre-defined parameter space at the expense of being
limited to the resolution and scope of the parameter space provided by
the authors.  For more generalized applications, SED synthesis 
packages are now available that offer a wider set of input parameters and 
the exploration of a continuous parameter space (e.g. stellar 
population synthesis codes such as GALAXEV, \citet{bruzual03} and 
pan-chromatic galaxy synthesis codes like GRASIL, \citet{silva98} and 
CIGALE, \citet{noll09}). Of course, one may use these packages to 
construct ones own template libraries \citep[e.g.][]{michalowski10} 
for specific applications as well.  In both cases, as the generality 
of the underlying physics increases to provide relevance to a wider 
class of sources, so does the number of free parameters in the models.  
The unavoidable existence of correlations in these parameters insists
that simply finding a best-fit parametrization is no longer
sufficient.  Rather, we now require tools that properly reveal the
complexities and correlations in the adopted model parameter space. 
 
With this in mind, here we present the general purpose MCMC-based SED
fitting code SED Analysis Through Markov Chains or
SATMC\footnote{http://www.ascl.net/1309.005}. Monte
Carlo Markov Chain (MCMC) techniques are a set of methods based in
the Bayesian formalism which enable efficient sampling of
multi-dimensional parameter spaces in order to construct a
distribution proportional to the probability density distribution of
the input parameters, known as the {\it posterior parameter distribution} 
or simply the {\it posterior}.  The posterior identifies the nuances in
parameter space, including any possible correlations, making MCMCs
particularly useful for exploring SED models and libraries with high 
dimensionality.  MCMC based SED fitting codes are relatively new
but have been growing in popularity
\citep[e.g.][]{sajina06,acquaviva11,serra11,pirzkal12}. In addition to
deriving the posterior for best fit parameter and confidence level
estimation, SATMC includes many features to aid in improving
performance and allows users to easily incorporate additional
knowledge and constraints on parameter space in the form of priors.
Additionally, SATMC versions exist in both IDL and Python and both are
modular and straightforward to use in any wavelength regime and for
any class of sources.

This paper is organized as follows.  We start by detailing the MCMC
algorithm and the basic process for MCMC-based SED fitting.  We then
provide case examples displaying the versatility and accuracy of SATMC
compared to standard least-squares methods.  As part of this
demonstration, we also present a set of SEDs derived from SATMC when
used in conjunction with the SED synthesis routine GRASIL to a sample
of X-ray selected starburst galaxies.  These fits highlight the key 
advantages obtained from the MCMC-based approach and their agreement 
with similar classes of composite SEDs. 

\section{SATMC: The MCMC Algorithm}

The primary motivation for SATMC is to provide a means for efficient
sampling of a parameter space with $n_D$ free parameters in order to derive
parameter estimates and their associated confidence intervals.  This
is accomplished by sampling an $n_D$-dimensional surface proportional to
the probability density function of the parameters given the data
$P(\textbf{x}|\textbf{D})$, also referred to as the {\it posterior}
parameter distribution.  We determine the posterior through Bayes Theorem
\begin{equation}
P(\textbf{x}|\textbf{D}) \propto P(\textbf{D}|\textbf{x})P(\textbf{x})
\end{equation}
where $P(\textbf{x})$ represents our current knowledge of the
parameters or {\it priors} and $P(\textbf{D}|\textbf{x})$ is the
probability of the data given the model parameters, often
referred to as the likelihood $L$.  We shall define the general form
of the likelihood according to
\begin{equation}
P(\textbf{D}|\textbf{x})=L=\prod^M_{i-1}{\rm exp}(\frac{(D_i-f(d_i|\textbf{x}))^2}{2\sigma^2_i})
\end{equation}
where the product runs over the M observed data points ${\bf D}$ whose
individual variances are given by $\sigma_i^2$ and the function 
$f({\bf d}|{\bf x})$ represents the model observations ${\bf d}$ for the 
set of ${\bf x}$ parameters.  Note, however, that this prescription of $L$
assumes that the observations have Gaussian distributed errors, an 
assumption we will return to later in \S~3.1.  In the following sections, 
we detail specific MCMC features that define the basic operation of SATMC. 

\subsection{MCMC Acceptance and Convergence}

Within the literature, there are a variety of sampling algorithms one may 
use to construct an MCMC \citep[e.g. Metropolis-Hastings or Gibbs
  sampling;][]{metropolis53,hastings70,geman84,geyer92,chib01,verde03,mackay03}.
For SATMC, we employ the Metropolis-Hastings algorithm which works as
follows: 
\begin{itemize}
\item Generate a proposal distribution $q({\bf x_i|x_{i-1}})$ from which 
the candidate steps ${\bf x_i}$ will be drawn
\item Calculate the acceptance probability according to the
likelihood ratio
($\alpha=$min(1,$\frac{P({\bf x_i}|{\bf D})q({\bf x_i|x_{i-1}})}{P({\bf x_{i-1}}|{\bf D})q({\bf x_{i-1}|x_i})}$)) 
\item Draw a uniformly distributed random number $u$ from 0 to 1 and
accept the step if $u<\alpha$, reject otherwise
\item Repeat for the next step
\end{itemize}

Though the exact choice of $q({\bf x_i|x_{i-1}})$ is
arbitrary, it is common to adopt an $n_D$-dimensional multivariate-Normal
distribution $\mathcal{N}$(${\bf \mu},\Sigma$) with
$\Sigma=\sigma_{\bf x}^2${\bf I} where {\bf I} is the identity matrix,
$\sigma_{\bf x}$ is the variance of each parameter in {\bf x} and
\begin{equation}
\mathcal{N}({\bf \mu},\Sigma)\propto |\Sigma|^{-\frac{1}{2}}{\rm
  e}^{-\frac{1}{2}({\bf x}-{\bf \mu})^{\prime}\Sigma^{-1}({\bf x}-{\bf \mu})}
\end{equation}
for generating proposed steps
\cite[e.g.][]{gelman95,roberts97,roberts01,atchade05}. Following these
works, it is found that the distribution $\mathcal{N}$(${\bf
  \mu},\Sigma$) should be tuned for optimal performance between 
the time necessary to reach a stable solution (referred to as the 
{\it burn-in} period) and sampling of the posterior, which may be
achieved by adjusting $\mathcal{N}$(${\bf \mu},\Sigma$) until the
resulting chains have acceptance rates of $\sim$23 percent in the limit
of large dimensionality.  We therefore base our proposal distribution 
around an adaptive covariance matrix $\Sigma$ 
\citep[similar to][]{acquaviva11} such that samples are drawn from the 
multivariate-Normal distribution now given by $\mathcal{N}$(${\bf
  x_{i-1}},\Sigma$).  To obtain an acceptance rate of $\sim$23
percent, we start by initializing the covariance matrix for each chain
as $\sigma_{\bf x}^2${\bf I} where $\sigma_{\bf x}$ is set
proportional to the input parameter ranges.  After a period of steps,
we then calculate $\Sigma$ directly from the chain over the previous
interval, i.e.  
\begin{equation}
\Sigma={\rm E}[({\bf x}-{\rm E}[{\bf x}])({\bf x}-{\rm E}({\bf x}))^T]
\end{equation}
where $E[{\bf x}]$ is the expectation value or weighted average of
${\bf x}$. Following each period of steps, we compute the acceptance
rate and scale $\Sigma$ if the acceptance rate is too high ($>$26
percent) or too low ($<$20 percent).  The covariance matrix is
continuously updated until the target $\sim$23 percent
acceptance is reached, which allows $\Sigma$ to take on the shape of
the underlying posterior to readily identify and account for possible
correlations in the parameters.   

Once the target acceptance rate has been reached, we then check for
chain convergence to determine when  the {\it burn-in} period is
complete.  In the MCMC literature there are many approaches to
determine convergence \citep[e.g.][]{gelman92,gilks96,raftery92}; we
opt for the Geweke diagnostic \citep{geweke92} which compares the
average and variance of  samples obtained in the first 10 percent and
last 50 percent of a chain segment.  If the two averages are equal
(within the tolerance set by their variances), then the chain is
deemed to be stationary and convergence is complete.  We check both
the acceptance rate and convergence, verifying the acceptance rate
before checking for convergence, over a default period of 1000 steps
until both have been satisfied.  If necessary, the covariance matrix
is modified to maintain the target acceptance rate. Once both criteria
are fulfilled, {\it burn-in} is completed and the chain(s) are set to
continue to provide sampling of the posterior. 
   
\subsection{Parallel Tempering}

\begin{figure}
\includegraphics[width=0.5\textwidth]{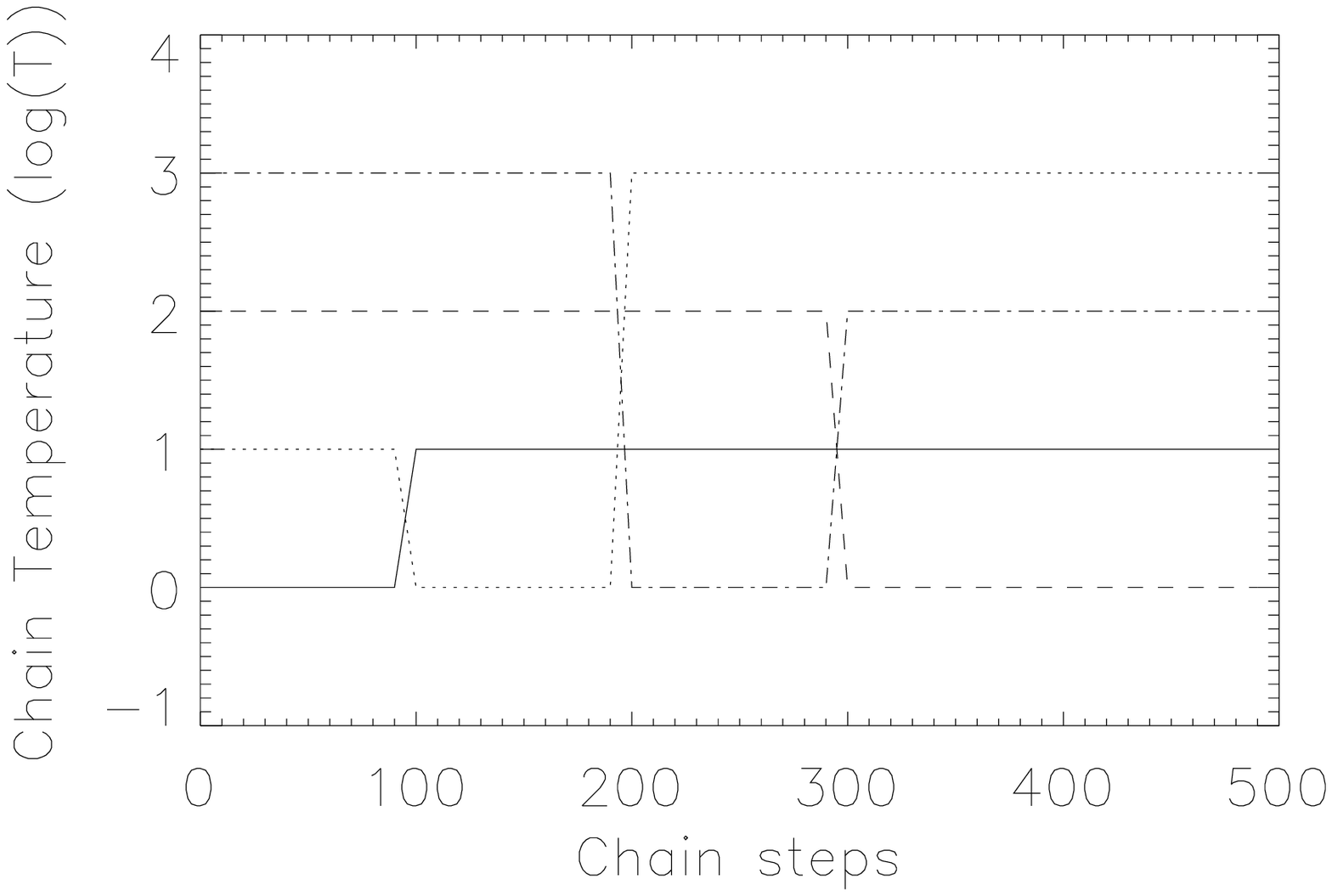}
\includegraphics[width=0.5\textwidth]{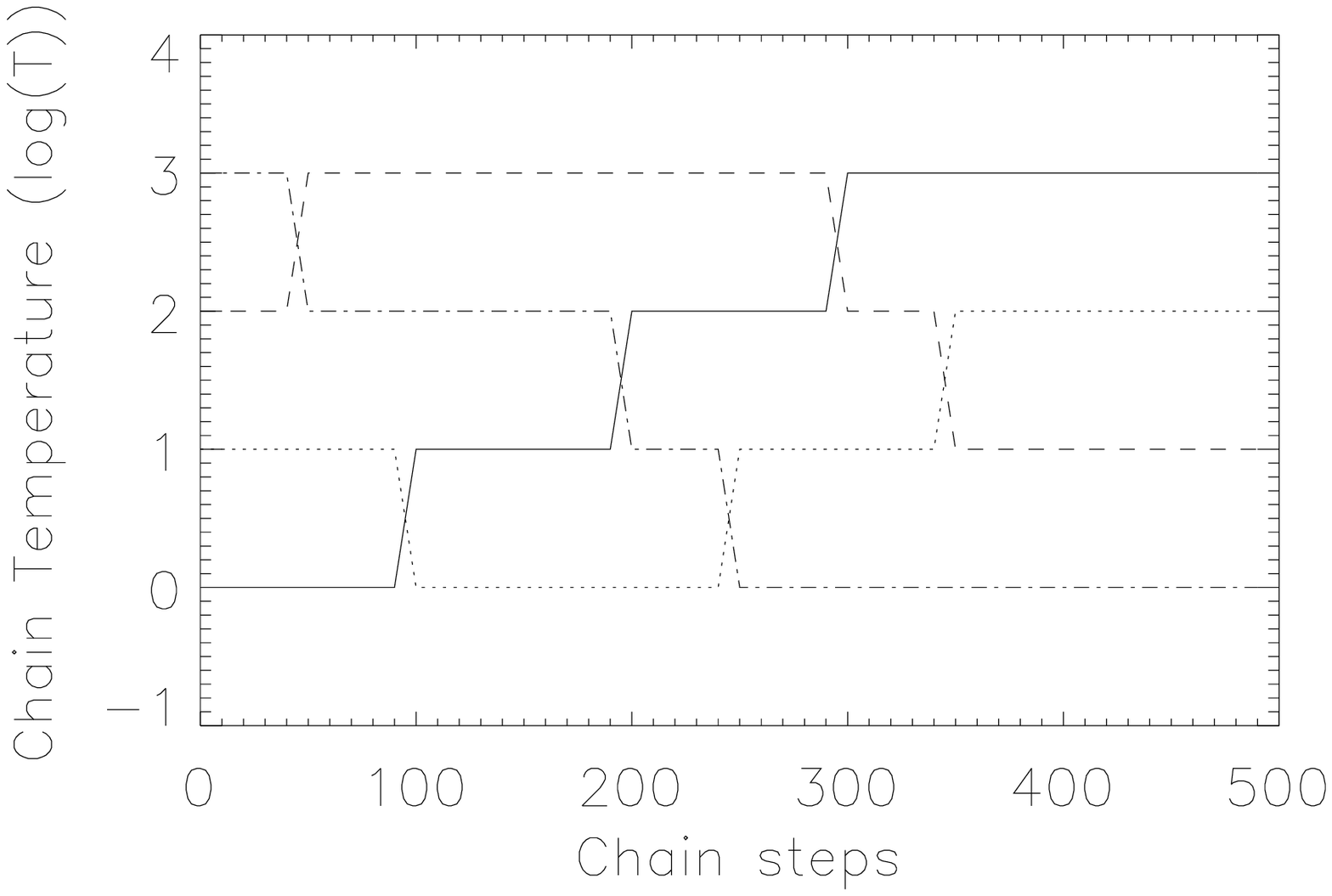}
\caption{Simple visualization of transitions between tempered
  chains. Transitions may occur either between the cold and tempered
  chains (top) or between adjacent pairs (bottom).  The former allows
  for rapid sampling of parameter space to quickly reach a global
  maxima while the latter improves sampling around a single location
  to avoid local maxima in the construction the posterior parameter
  distribution.  In this example, chain transitions are proposed after
  every 100 steps; in practice, these transitions are proposed every
  3000 steps.}
\label{fig:tempedchains}
\end{figure}

When constructing the posterior distribution, we must be
careful to sample all of the nuances of parameter space as the
posterior distribution is not guaranteed to be a smooth or even
continuous function of the free parameters.  As with all fitting 
approaches, the existence of {\it local} maxima in the posterior requires
that either we known {\it a priori} the general location of the global
maximum or we implement a sampling technique capable of increasing the 
probability of finding it.  For example, one could choose
to initialize a large number of chains that cover random locations
throughout parameter space.  While this approach is nearly guaranteed to
find the global maximum, it is also extremely inefficient.  SATMC utilizes 
a technique known as ``Parallel Tempering'' \citep[see review by][]{earl05} 
which, in analogy to simulated annealing, uses several chains -- each 
with progressive modifications to likelihood space parametrized as a 
statistical 'temperature' --  to search parameter space and exchange 
information about the posterior at each chain's location.  For a given
chain at temperature T, the likelihood is 'flattened' according to 
$L_T=L^{1/(1+T)}$. The tempered chains thus distort likelihood 
  space, exchanging sensitivity to the details within posterior for the 
  general shape, such that chains with higher temperatures will accept more 
steps, and thus sample larger regions of parameter space.  By coupling
these tempered chains, we allow 'colder' chains to access areas of
parameter space they may have otherwise been unable to reach.  The
process for handling tempered chains works as follows:  
\begin{itemize}
\item Temperatures are assigned to chains in a progressive manner 
  with one chain designated the fiducial 'cold' chain with $T=0$
  (e.g. $T=[0,10,100,1000]$).
\item Chains are allowed to progress for a set number of steps
  (SATMC uses 3 iterations of acceptance/convergence
  or 3000 steps by default)
\item A 'swap' of chain state information is proposed using the 
  Metropolis acceptance algorithm 
 $\alpha={\rm
    min}(1,\frac{(P({\bf x_i}|{\bf D}))^{1/(1+T_j)}(P({\bf x_j}|{\bf D}))^{1/(1+T_i)}}{(P({\bf x_i}|{\bf D}))^{1/(1+T_i)}(P({\bf x_j}|{\bf D}))^{1/(1+T_j)}}$
\item The new set of chains is then allowed to run until the next swap
  of chain state information.
\end{itemize}
This process of coupling individual chains with a Metropolis-Hastings
acceptance algorithm was initially proposed by \cite{geyer91} and is
referred to as a Metropolis Coupled Monte Carlo Markov Chain or
MC$^3$.   When applying temperatures to individual chains, the 
sampling of each chain will vary due to the distortions of likelihood
space; i.e., a chain at a higher temperature will accept more steps than
a chain with an identical proposal distribution at lower temperature.  To 
maintain a reasonable sampling of the tempered likelihoods, each chain 
is treated individually for the purposes of acceptance and convergence 
testing.  Note, however, that updating the tempered chains occurs only 
with the acceptance and convergence testing of the cold chain.  This 
ensures that we obtain a proper sampling of the posterior without 
being influenced by the otherwise distorted likelihood space.  In parallel 
tempering techniques, one can specify whether the potential swaps occur only 
between the cold and any tempered chain (0,$j$) or between adjacent 
chains ($i,i\pm$1) (see Fig.~\ref{fig:tempedchains}).  The former 
allows for rapid sampling and mixing of the chains to determine the 
global maximum and is thus used during the {\it burn-in} period.  
The latter passes state information down through the tempered chains 
so that the cold chain may access a more representative region of 
parameter space; this method is used after {\it burn-in} to properly 
sample parameter space around the maximum likelihood.  In this case, 
chains progress for a set length (500 steps by default) whereafter 
the chain transitions are proposed. Should a swap be made with the 
cold chain after {\it burn-in} is complete, we re-compute the 
acceptance rate and convergence as outlined in the previous section 
to verify proper sampling.  Due to the modified acceptance rate 
and misshapen likelihood space of the tempered chains, 
it is generally undesirable to use them in re-constructing the 
posterior parameter distribution.  If there have been no swaps to the 
cold chain after 10 iterations, the temperatures of all chains are set 
to 0 and the MCMC is allowed to continue until sufficient samples of the 
posterior distribution around the maximum have been obtained.  

\section{SATMC SED Specific Features}

The methods presented in \S~2 are generalized for any MCMC-based
algorithm.  Here, we detail specific modifications and methods
utilized in SATMC for SED fitting.

\subsection{Observations and Upper Limits}

In order to perform a fit, SATMC requires at least two sets of
information: 1) a file containing the M observations including their 
wavelengths ($\lambda_i$) or frequencies ($\nu_i$), observed fluxes 
($f_{\rm obs,i}$) and corresponding uncertainties ($\sigma_{\rm obs,i}$) and 
2) the model libraries. Following from Eqn.~2, we define the likelihood 
for a given set of observations and models as 
\begin{equation}
L=\displaystyle\prod^M_{i-1} {\rm exp}(-\frac{(f_{\rm obs,i}-f_{\rm
    mod,i})^2}{2\sigma_{f_{\rm obs,i}}^2})
\end{equation}
or similarly
\begin{equation}
\ln(L)=\displaystyle\sum^M_{i-1} -\frac{(f_{\rm obs,i}-f_{\rm
    mod,i})^2}{2\sigma_{f_{\rm obs,i}}^2}
\end{equation}
where $f_{\rm mod,i}$ are the model fluxes which for narrow passbands 
are interpolated model SED values at the observed
frequency/wavelength.  Alternatively, $f_{\rm mod,i}$ may be
calculated from a set of passbands with a given filter response
($p(\nu)$ as provided by the user) as
\begin{equation}
f_{mod,i}=\frac{\int_{\nu_{min}}^{\nu_{max}} {\rm SED}(\nu) p_i(\nu) {\rm~d}\nu}{\int_{\nu_{min}}^{\nu_{max}} p_i(\nu) {\rm~d}\nu}
\end{equation}
where $\nu_{min}$ and $\nu_{max}$ are taken from the filter files.

\begin{figure}
\includegraphics[width=0.45\textwidth]{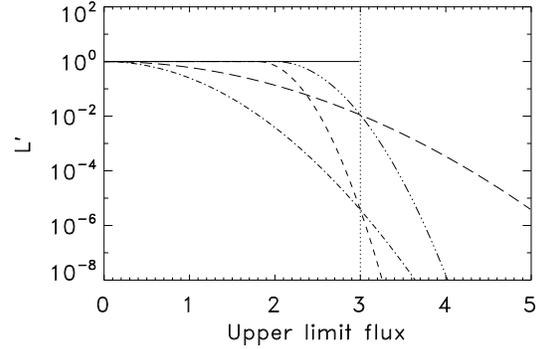}
\caption{Example likelihood distribution for upper limits.  Here, the
  upper limit has a value of 3 mJy (dotted vertical line).  The remaining
  lines represent the likelihood distribution for the simple step function
  (solid), one-sided Gaussians assuming the upper limit is at the 5$\sigma$
  level with a cut-off at 1.8 mJy (dashed) and 0 mJy (dot-dash) and one-sided
  Gaussians with the upper limit at the 3$\sigma$ level and cut-offs at
  2 mJy (triple dot-dash) and 0 (long dash). }
\label{fig:osgauss}
\end{figure}

In cases where there is no detection but only an upper limit, SATMC
provides a parametrized probability distribution to allow the user to
reflect their confidence in the underlying statistics of the
observations.  A simple approach for incorporating upper limits is to
assume a step function accepting all models that fall below the limit
and rejecting all that lie above.  While attractive in its simplicity,
this method harshly truncates regions of parameter space and
improperly weights the upper limit's contribution to $L$.  To
partially alleviate these issues, SATMC implements a one-sided
Gaussian distribution so that models with fluxes above the upper limit
may contribute to the posterior with a small but non-zero probability
\citep[see also][]{feigelson85,isobe86,sawicki12}.  The one-sided
Gaussian is defined by  
\begin{equation}
L'\propto\left\{ \begin{array}{cl}
 1 &\mbox{ if $f_{mod}< f_{\rm co}$} \\
 {\rm exp}(-\frac{(f_{\rm co}-f_{mod})^2}{2\sigma_{\rm co}^2}) &\mbox{ if $f_{mod}\ge f_{\rm co}$}
       \end{array} \right.
\end{equation}
where $f_{\rm co}$ represents the cut-off transition from flat $L'$=1
to a  Gaussian 'tail' with standard deviation $\sigma_{co}$.
Parametrizing the probability distribution in this manner allows for
a compromise between the simple step function, also available in
SATMC, and over-interpreting the shape of the noise distribution at
small fluxes, as demonstrated in Fig.~\ref{fig:osgauss}.

\subsection{Photometric Redshift Estimation}

Since redshift is just another parameter in a source's SED, SATMC is
capable of providing photometric redshift (photo-z) determinations for
sources in the context of the input SED models. Traditionally,
photo-z codes implement least-squares methods
\citep[e.g. HYPERZ ][]{bolzonella00} though some recent codes have
adopted the Bayesian formalism for using priors (e.g. BPZ,
\citealt{benitez00}; EAZY, \citet{brammer08}).  

The standard approach in photo-z estimation is to apply an SED library
(often limited to a few galaxy types) to find the photo-z and then
repeat the fitting using the same (or in some cases different) library
fixed at the photo-z to estimate source properties.  This approach
under-estimates the true errors on the photo-z and other fitted
parameters.  SATMC fits the redshift simultaneously with 
all other parameters and produces a direct determination of the
redshift-parameter probability distributions in addition to the
marginalized redshift probability distribution
$P(z)$\footnote{Cosmology is presently fixed in SATMC and assumes flat
$\Lambda$CDM with H$_0$=70, $\Omega_{\lambda}$=0.7 and $\Omega_M$=0.3.}.  

We have tested and improved the photo-z estimation with SATMC in
collaboration with the CANDELS team.  \cite{dahlen13} provides a
complete analysis of various photometric redshift estimation
techniques for samples of CANDLES galaxies in the GOODS-S field
\citep{giavalisco04}.  Out of the 13 participating groups, SATMC was 
the only MCMC-based code used to generate photo-z's.  Since the tests 
reported in Dahlen et al., we have reduced the outlier fraction
(fraction of sources with $(z_{spec}-z_{phot})/(1+z_{spec})>0.15$) from
$\sim$9-14 percent to $\sim$3-8 percent through modification of our
input templates, luminosity priors \citep[e.g.][]{kodama99,benitez00}
and zero-point photometry corrections \citep[e.g.][]{dahlen10}. 

\subsection{Template Libraries and SED Synthesis Routines}

As MCMC samplers require a continuous parameter space, SATMC allows
the user to incorporate SED synthesis routines (e.g. GALAXEV, GRASIL,
CIGALE) to generate SEDs at candidate steps and compute the resulting
likelihoods.  Empirical templates are often defined with a scalable
normalization factor as one of the few (if any) free parameters which
remains continuous when used with SATMC.  Template libraries,
unfortunately, rarely offer a fully continuous parameter space; often
mixing sets of continuous (e.g. model normalization) and discretely
sampled parameters.  To create a pseudo-continuous space from such
template libraries, SATMC computes the likelihoods of models
bracketing the current step according to Eqns.~5 \& 8 and applies
multi-linear interpolation to determine the likelihood of the current
proposed step.  We emphasize that SATMC may be used for \textit{any}
class of SED models (empirical, template library or synthesis
routine), so long as the appropriate interface is constructed by the
user.  One should note, however, that there is a trade-off between the
discretely sampled template libraries or empirical templates and
continuous parameter space offered by SED synthesis routines \citep[see
  also][]{acquaviva11,acquaviva12} since the run-time of the SED
synthesis routines will dominate the MCMC calculation (e.g. 3+ days
run-time with GRASIL versus 3-5 minutes with a template library). 

Regardless of which class of SED models one wishes to adopt, empirical 
and theoretical SED models are commonly derived for a particular 
physical process and/or over a particular wavelength regime
\citep[e.g.][]{bruzual93,efstathiou00,siebenmorgen04}.  To fully
reproduce a galaxy's SED, additional components may be required either
to complete the wavelength coverage or to include a missing physical
process (e.g. adding AGN emission to a star-formation template set).  
SATMC will construct a linear combination of multiple input SED models
under the assumption that the underlying physical processes are
independent.  The MCMC process itself does not change: the combination 
of two models with $N1$ and $N2$ free parameters, respectively, is 
viewed as a single model with $N1+N2$ parameters when calculating 
likelihoods and taking potential steps. 

\subsection{Inclusion of Priors}

An added feature of SATMC is the ability to include additional
information to provide additional weights and constraints to
likelihood space.  In the Bayesian formalism, this extra information forms
the priors of Eqn~1.  For our implementation, we expand the definition
of priors from the traditional Bayesian definition to include options
for limiting and inherently correlating parameter space.  This was
deemed necessary for circumstances of fitting multiple template
libraries where parameters from each model have the same physical
interpretation and thus are not independent (e.g. A$_V$ from one model
library and optical depth in another) and cases where additional
information not available to the fits is available (e.g. restricting
the age of a galaxy at a given redshift).

\subsection{Application of the Posterior}

A final feature of SATMC lies in the determination of the posterior
parameter distribution.  As we store the likelihood and location in
parameter space for each step, it becomes a simple task to construct
parameter confidence intervals and even parameter-parameter confidence
contours to examine relative parameter degeneracies
(see Figure~\ref{fig:sk07}).  Unfortunately, in order to visualize
an $n_D$-dimensional parameter space, we must project or 'marginalize'
parameter space into a one or two dimensional form.  When
marginalizing sets of parameters, the true shape of the posterior will
be distorted which may not reveal correlations in higher dimensions.
This also leads to a simplification when reporting the confidence
intervals on individual parameters as traditional terms such as
'1$\sigma$' imply Gaussianity in the posterior which is unlikely to
exist.  Instead, one dimensional confidence levels are 
produced from the parameter range where 68 percent of all accepted
steps are contained, marginalized over all other free parameters.
Throughout the text, 'errors' quoted when derived from SATMC refer to
these marginalized parameter ranges.

\section{Testing of the MCMC Algorithm}

\begin{table*}
\caption{Comparison of best-fit \citealt{sieben07} parameters using
standard techniques (SK07) and SATMC.  For M82, we add an unreddened
blackbody of temperature $T=2500$ K to the SK07 library for
consistency.  Note also that SK07 suggest two models for Arp220, one
with $R=1$ and $A_V=120$ and another with $R=3$ and $A_V=72$ (the
latter being the accepted best-fit template), which are represented in
our fitting through the large uncertainties in $R$ and $A_V$.  For
each model, we also report the log-likelihood (ln(L)) of the fit. Errors
are estimated at the 68 percent confidence level.}
\label{tab:compare}
\begin{tabular}{lccccccccccc}
Name & \multicolumn{5}{c}{SK07} & \multicolumn{6}{c}{SATMC} \\
 & $L_{tot}$ & $R$ & $A_{V}$ & $L_{OB}/L_{tot}$ & $n$ & $L_{tot}$ &
 $R$ & $A_{V}$ & $L_{OB}/L_{tot}$ & $n$ & ln(L)\\ 
\hline
M82    & 10$^{10.5}$ & 0.35 & 36 & 0.4 & 10000 & 10$^{10.5^{+0.2}_{-0.1}}$ & 0.38$^{+0.57}_{-0.03}$ & 35.2$^{+33.9}_{-14.8}$ & 0.418$^{+0.167}_{-0.018}$ & 9710$^{+290}_{-3907}$ & -1218.7\\
Arp 220 & 10$^{12.1}$ & 3    & 72 & 0.4 & 10000 & 10$^{12.1^{+0.1}_{-0.2}}$ & 2.84$^{+0.16}_{-1.70}$ & 70.1$^{+69.5}_{-32.2}$ & 0.410$^{+0.176}_{-0.010}$ & 190$^{+6883}_{-90}$ & -132.4\\
\end{tabular}
\end{table*}

\begin{figure*}
\hbox{\subfigure{\includegraphics[width=0.5\textwidth]{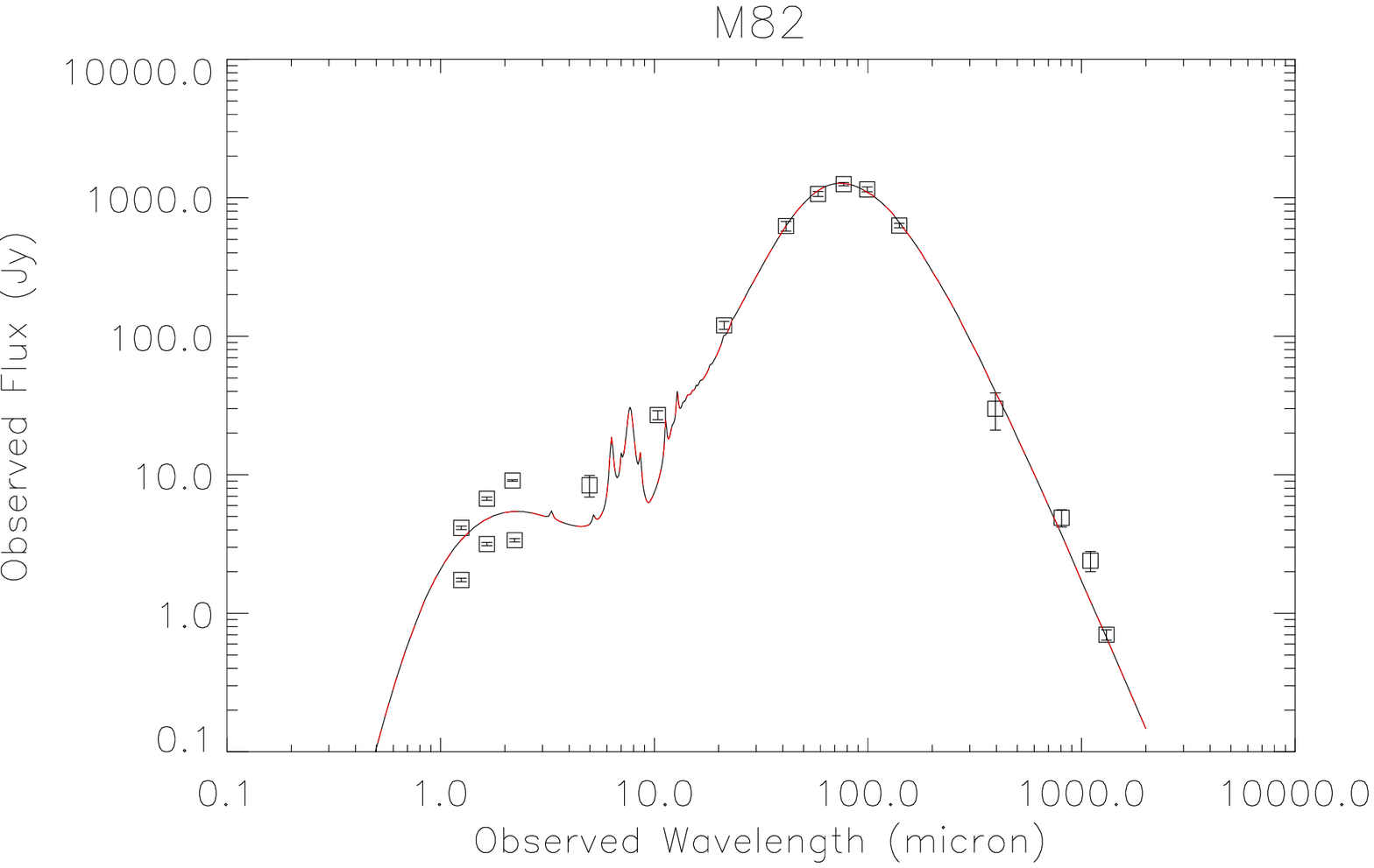}}
\subfigure{\includegraphics[width=0.5\textwidth]{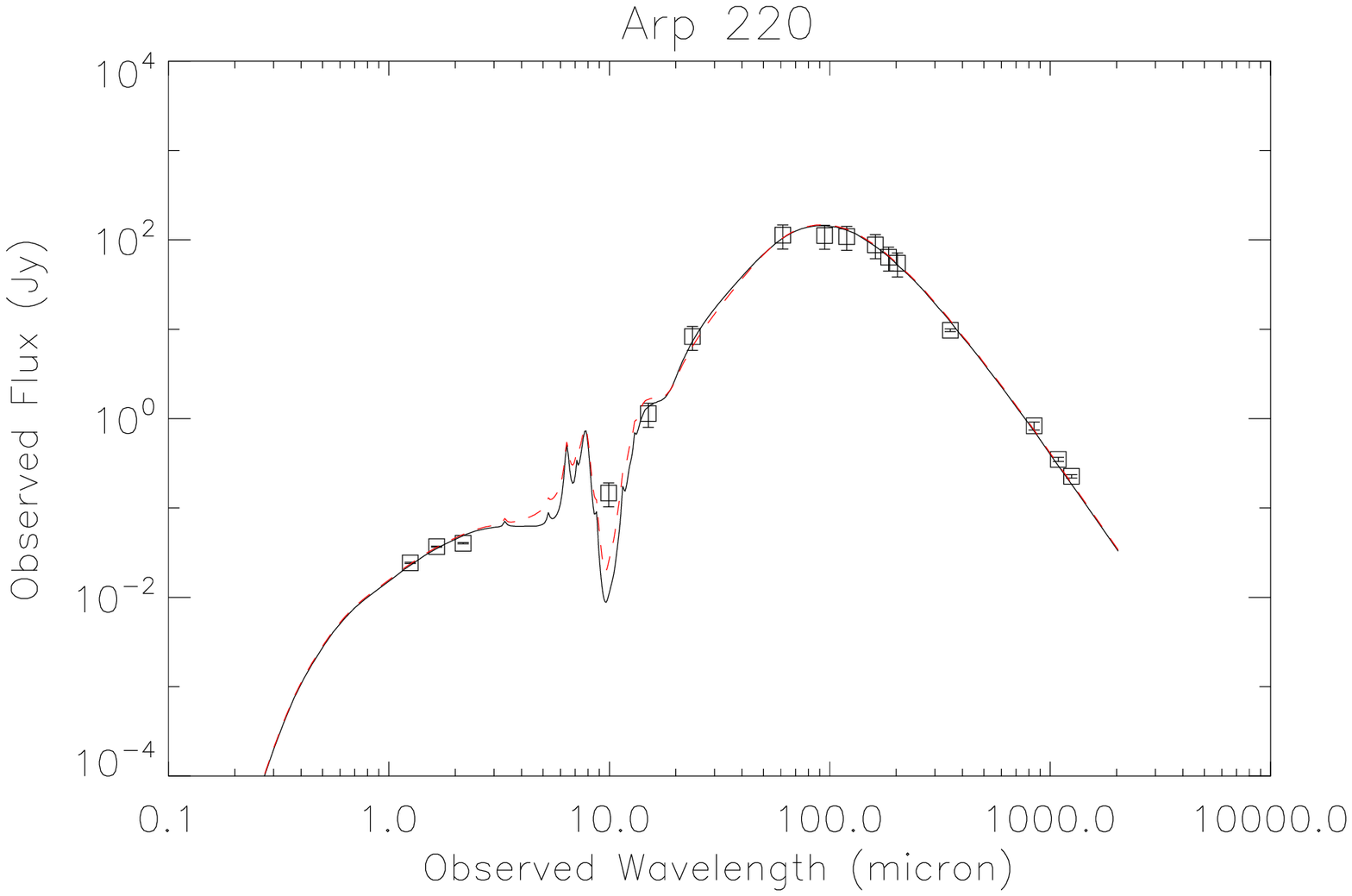}}}
\hbox{\subfigure{\includegraphics[width=0.5\textwidth]{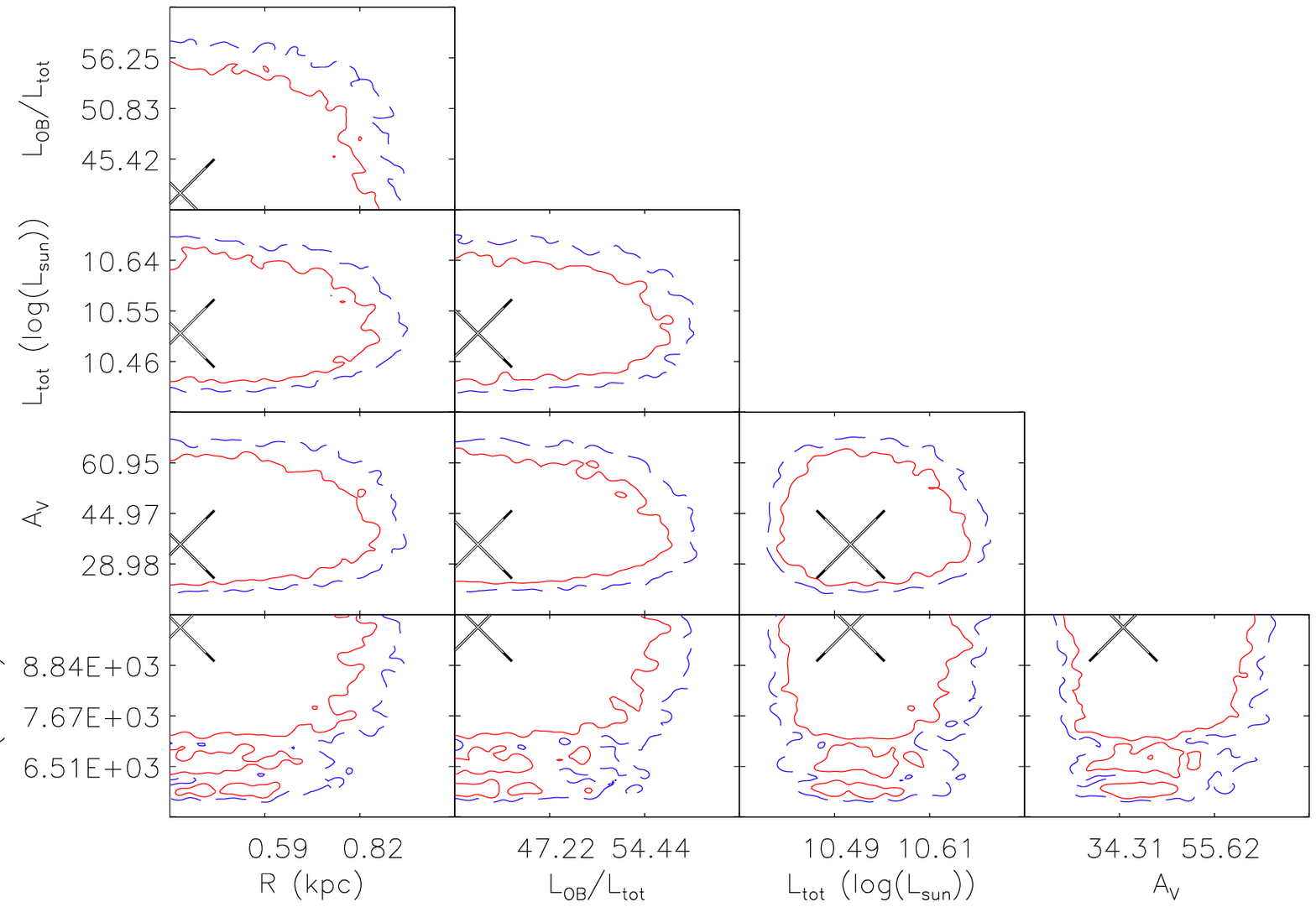}}
\subfigure{\includegraphics[width=0.5\textwidth]{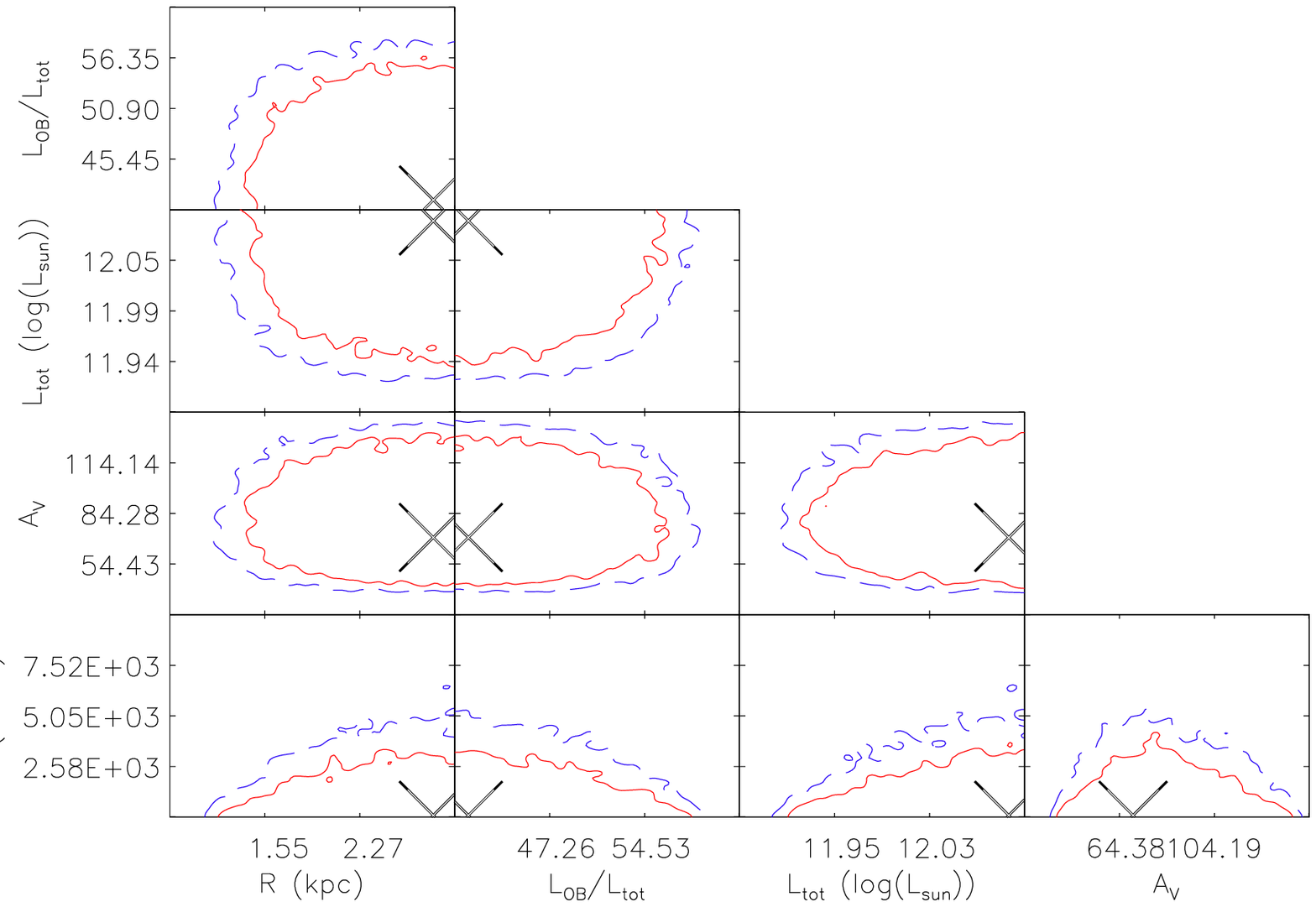}}}
\caption{Best fit SEDs for M82 (left) and Arp 220 (right) using the 
  SK07 template library. {\it Top:}  Observed photometry along with the
  best fit models from SK07 (dashed red line) and those obtained with 
  SATMC (solid black).  An unreddened blackbody of temperature $T=2500$ K
  has been added to M82 templates for consistency with SK07. {\it
    Bottom:} Smoothed parameter-parameter likelihood surfaces.  The 
  location of the maximum likelihood is marked by the large 'X'.  Contours 
  are placed at $\Delta$ln(L) intervals corresponding to the 68 and 90 
  percent confidence contours. The contours show truncation due to the
  parameter limits of the SK07 template library.} 
\label{fig:sk07}
\end{figure*}

With the SATMC algorithm as outlined in \S~2 and 3, we now set out to
verify the fitting results by analyzing sets of well known sources and
template libraries.  Though the examples provided here are for galaxy
SED modeling, SATMC makes no distinction between source types and may
just as easily be applied to Galactic sources, localized regions
within a particular galaxy or spectroscopy of individual sources.  We
begin with the well known galaxies Arp 220 and M82 and the starburst
SED library of \cite{sieben07} (hereafter SK07) which has already been
shown to provide reasonable fits to Arp 220 and M82.  The SED library
consists of over 7000 templates with emission from a starburst
parametrized by its total luminosity $L^{tot}$, nuclear radius $R$,
visual extinction $A_V$, ratio of luminosity from OB stars to the
total luminosity and hot spot dust density $n$.  The observed SEDs for
Arp220 and M82 were constructed from data available on the NASA/IPAC
Extragalactic Database (NED\footnote{http://ned.ipac.caltech.edu/})
over the 1-1500 $\mu$m wavelength range.  To ensure a meaningful
comparison to SK07, we utilize the same photometric data for M82 and
Arp 220 (including multiple aperture JHK photometry for M82) such that
the only difference lies in the fitting method.
Table~\ref{tab:compare} provides the best fit models as obtained by
SK07 and SATMC with the models and parameter-parameter confidence
contours shown in Figure~\ref{fig:sk07}.  Note that while SATMC will 
sample all of parameter space within that defined by the input template 
library, it is not possible to extrapolate likelihood information 
beyond the limits of the templates.  This effect is responsible for the 
apparent truncation of parameter space seen in Figure~\ref{fig:sk07}.  
Despite the irregular, non-uniform parameter sampling of the SK07 
templates, SATMC closely recovers similar values to SK07 for the best 
fit model parameters.  The parameter-parameter constraints as shown in
Figure~\ref{fig:sk07} highlight the uncertainty in applying
template sets; particularly for Arp 220 where SK07 suggest two likely
best-fit templates.  

\begin{figure*}
\hbox{\subfigure{\includegraphics[width=0.5\textwidth]{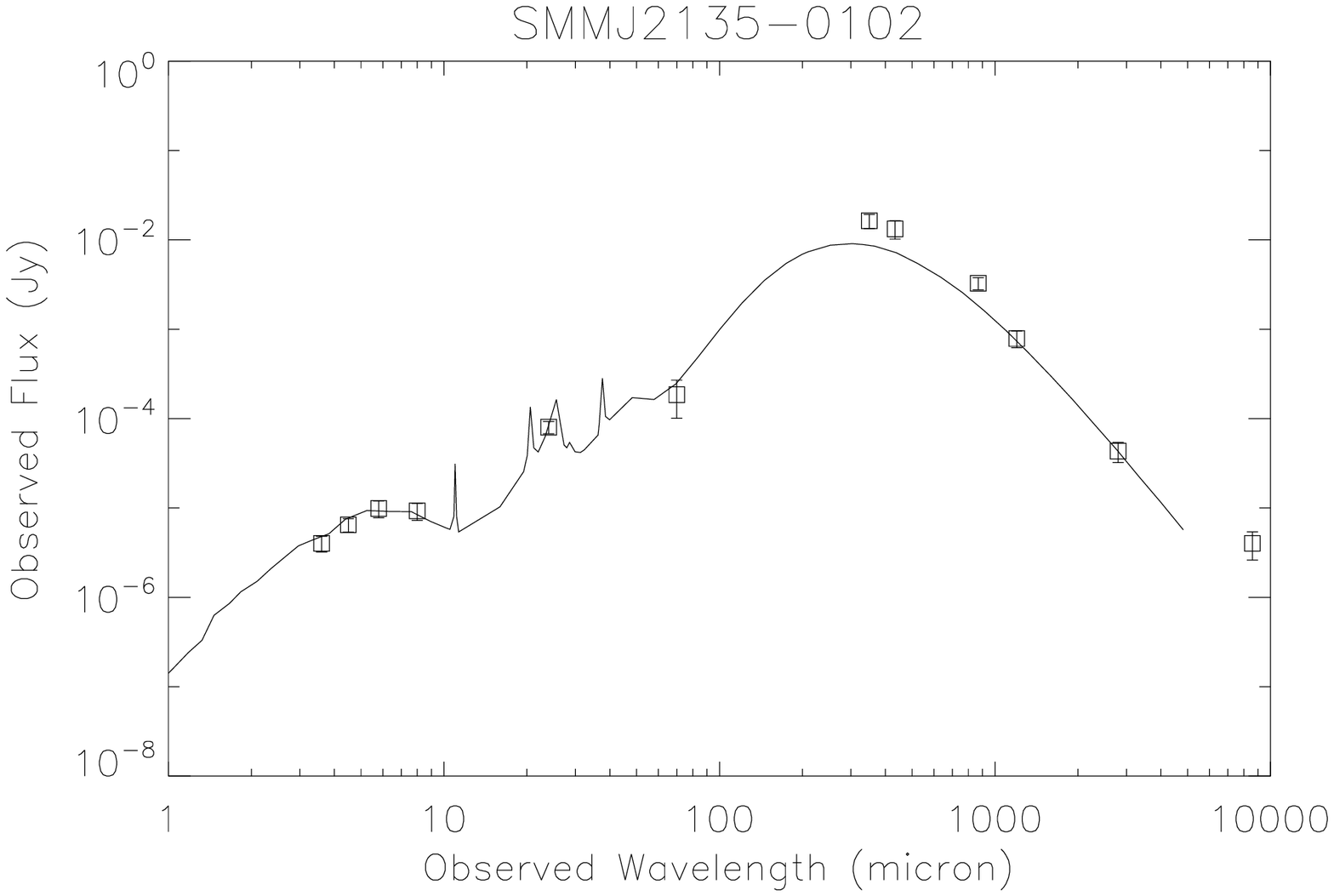}}
\subfigure{\includegraphics[width=0.5\textwidth]{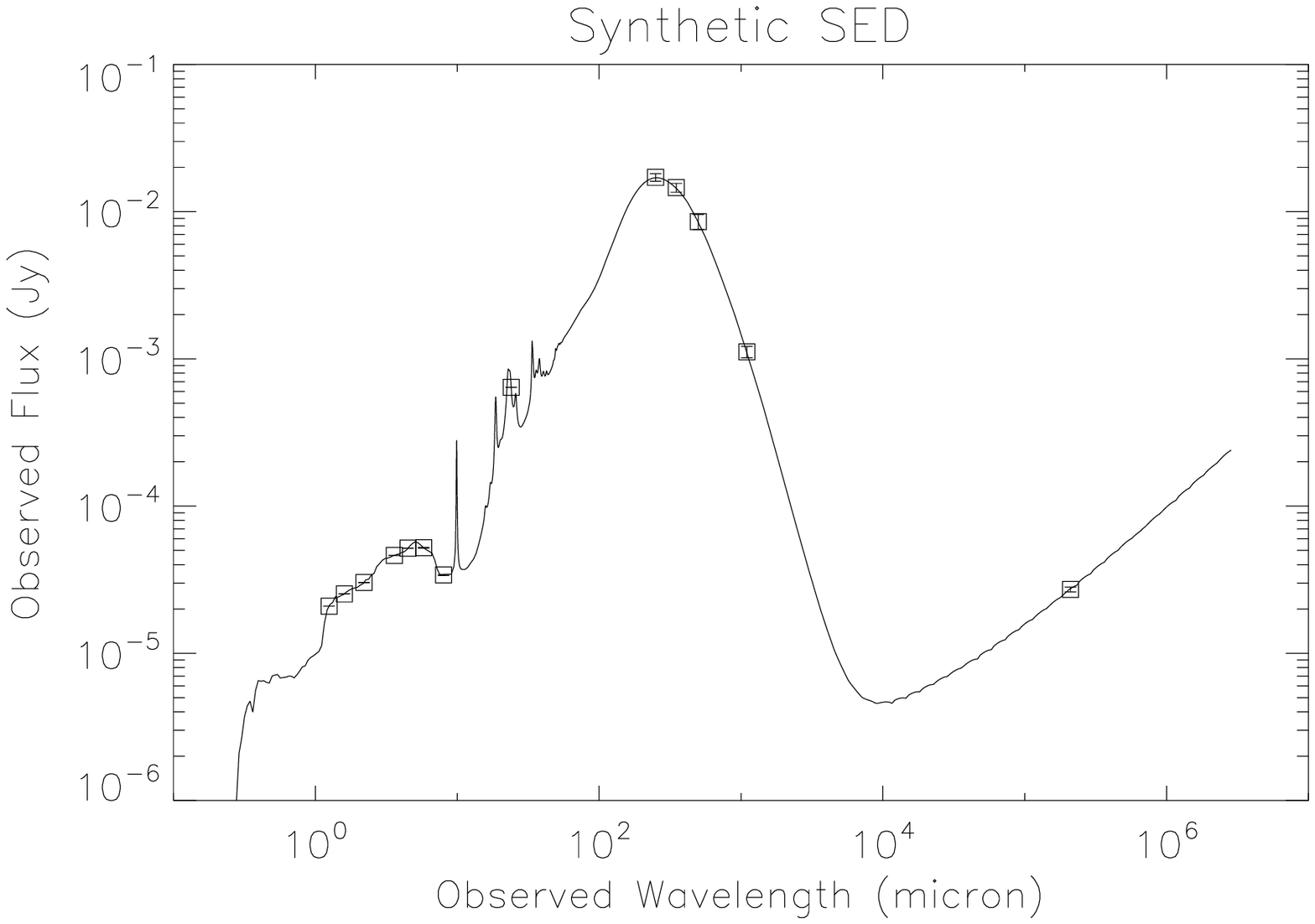}}}
\caption{{\it Left:} Best fit ERS00 model to the observed SED of 
  SMMJ2135-0102. {\it Right:} Best fit GRASIL model to our synthetic $z=2$
  SED.}
\label{fig:sedfits}
\end{figure*}

For a more realistic test in determining the physical properties of a
galaxy, we turn to the bright, lensed submillimetre galaxy SMM
SMMJ2135-0102 \citep{swinbank10} and apply the template SED library of
\cite{efstathiou00} (hereafter ERS00).  The ERS00 starburst library
consists of 44 templates with emission parametrized simply by the age
of the starburst and the optical depth of molecular clouds where the
new stars are forming.  A normalization factor is required with the
ERS00 templates to scale the emission from a single giant molecular
cloud (on which the templates were formulated) to the entire system.
This normalization factor roughly translates into a SFR with the model
age according to
\begin{equation*}
SFR\approx Norm*\rm{e}^{-t/20\rm~Myr}
\end{equation*}
as ERS00 assumes an exponentially decaying SFH with an e-folding time
of 20 Myr.  Applying the templates to the lensing-corrected SED of
SMMJ2135-0102, we find best-fit parameters of $1960^{+282}_{-250}$,
$56.6^{+10.5}_{-15.2}$ Myr and 199.9$^{+0.1}_{-44.4}$ for model
normalization, age and optical depth, respectively; this model is
shown in the left panel of Figure~\ref{fig:sedfits}.  Using the
standard FIR-SFR relation of \cite{kennicutt98}, the best-fit
parameters imply a SFR of
$\sim$192$^{+28}_{-25}\rm~M_{\odot}~yr^{-1}$.  Comparatively, Swinbank
et al. fit SMMJ2135-0102 with a two temperature modified blackbody and
derived a SFR of 210$\pm$50 M$_{\odot}$ yr$^{-1}$ using \cite{kennicutt98};
fully consistent with our results.   

\begin{table}
\caption{Comparison of input and recovered GRASIL parameters with
SATMC.  Errors are estimated at the 68 percent confidence level.} 
\label{tab:simsed}
\begin{tabular}{lp{2cm}cc}
Parameter & Description & Input & Best-fit values \\
\hline
T$_{\rm gal}$           & Galaxy age (Gyr) & 2.0 & 2.00$^{+0.01}_{-0.01}$\\
CLOUD\_RATIO       & Mass/Radius$^2$ for molecular clouds M$_{\odot}$ pc$^{-2}$ & 1000  & 990.6$^{+33.5}_{-35.7}$\\
M$_{\rm final}$         & Galaxy mass accumulated over T$_{gal}$ (M$_{\odot}$) & 1e12 & 9.99$^{+0.16}_{-0.10}$e11\\
M$_{\rm burst}$         & Mass of stars formed during starburst (M$_{\odot}$) & 1e10 & 1.02$^{+0.01}_{-0.04}$e10\\
M$_{\rm dust}$          & Dust mass (M$_{\odot}$) & 1e9 & 9.98$^{+0.21}_{-0.30}$e8\\
$\nu_{\rm sch}$         & Efficiency of Schmidt SFR & 0.3 & 0.300$^{+0.002}_{-0.006}$\\
ln(L)              & ln(L) of fit  & &-0.01 \\
\end{tabular}
\end{table}

As a final verification of SATMC, we create a simulated SED
with the SED synthesis routine GRASIL and attempt to recover the input
parameters.  GRASIL allows for numerous different parameters to be
specified that will then determine the physical scale, chemical
evolution and various attributes including dust content and inclusion
of a starburst component.  For a full description of GRASIL and its
parameters, we refer the reader to \cite{silva98}.  For our synthetic
galaxy, we set the galaxy age and total mass to 2 Gyr and 10$^{12}$
M$_{\odot}$, respectively.  A moderate amount of dust was included in
the form of a dust mass of 10$^{9}$ M$_{\odot}$.  The optical depth of
UV/optical photons depends on the ratio of the mass of the molecular
clouds in which new stars are formed and their size; we parametrize
this as the 'CLOUD\_RATIO' with an input value of 1000 M$_{\odot}$
pc$^{-2}$.  We then allow the simulated SED to undergo a 'merger' at 1.95
Gyr.  In effect, we set two SFH, one where the simulated galaxy is
following a standard Schmidt SFR (SFR=$\nu_{sch}$M$_{gas}^k$ where k
is fixed at 1 and $\nu_{sch}$=0.3) and one for the 'merger-triggered'
starburst which will convert 10$^{10}$ M$_{\odot}$ of gas into stars
over a 50 Myr period following an exponentially decaying SFH with
e-folding time of 50 Myr.  The two SFHs are then combined before
producing the final SED.  Table~\ref{tab:simsed} provides a
summary of our adopted parameters.  The resulting GRASIL spectrum was
then redshifted to $z=2$ and sampled at simulated wavelengths of
1.25$\mu$m, 1.6$\mu$m, 2.2$\mu$m, 3.6$\mu$m, 4.5$\mu$m, 5.8$\mu$m,
8.0$\mu$m, 24$\mu$m, 250$\mu$m, 350$\mu$m, 500$\mu$m, 1.1mm and 21cm
corresponding to the wavelength coverage of the near-IR JHK bands,
\textit{Spitzer}/IRAC+MIPS, \textit{Herschel}, AzTEC 1.1mm and VLA
observations.  Errors in each of these bands were chosen to be
representative of similar observations, from $\sim$0.01-0.1 $\mu$Jy in
the near/mid-IR and radio bands to $\sim$0.1-1.0 mJy for the (sub)mm.
A small amount of random noise generated from the uncertainty in each
band was added to the simulated data points prior to fitting.  The
fitting process for the simulated SED is then the same as previous
cases though we now use GRASIL to produce SEDs rather than refer
to a template library.  The results of our fitting are shown in the
bottom right panel of Figure~\ref{fig:sedfits} and
Table~\ref{tab:simsed}.  In addition to the synthetic GRASIL template
shown here, we have also tested SATMC using several different
combinations of input parameters and error handling (e.g. errors
proportional to the simulated observations or representative errors
as used previously) which all show similar agreement between the input
and fitted SEDs.  The excellent agreement between standard
least-squares fits, our synthetic input SEDs, and the recovered
parameters with SATMC shows that our MCMC-based SED fitting method is
robust and reliable.  

\section{Application of SATMC to SMGs}

\begin{table*}
  \caption{Counterpart IDs and redshifts to the X-ray-detected SMGs in
  our sample.  Counterparts are referenced according to their IAU
  identifier or catalog ID number if not available.  Redshifts given
  are the adopted spectroscopic/photometric values from \citealt{xue11}
  with photometric redshifts in parentheses.}
  \label{tab:sample}
  \begin{tabular}{@{}llllllr}
    \hline
    SMM ID & \textit{Chandra} ID & HST ID & \textit{Spitzer} ID &
    \textit{Herschel} ID & VLA ID & Redshift\\
    \hline
    AzGS7b & J033205.34-274644.0 & J033205.35-27454431 & J033205.35-244644.07 & Herschel72  &     & (2.808)\\
    AzGS10 & J033207.12-275128.6 &                     & J033207.09-275128.96 & Herschel435 &     & (1.829)\\
    AzGS1  & J033211.39-275213.7 &                     & J033211.36-245213.01 & Herschel280 & 660 & (3.999)\\
    AzGS13 & J033212.23-274620.9 & J033212.22-274620.7 & J033212.23-274620.66 & Herschel375 &     & 1.034\\
    AzGS7  & J033213.88-275600.2 &                     & J033213.85-275559.93 & Herschel1196& 336 & (6.071)\\
    AzGS11 & J033215.32-275037.6 & J033215.27-275039.4 & J033215.30-275038.31 & Herschel353 & 390 & (0.682)\\
    AzGS17a& J033222.17-274811.6 & J033222.14-244811.3 & J033222.16-274811.35 & Herschel1316&     & (1.760)\\
    AzGS17b& J033222.56-274815.0 & J033222.57-274814.8 & J033222.54-274814.94 & Herschel1316&     & (2.535)\\
    AzGS16 & J033238.01-274401.2 & J033238.04-274403.0 & J033238.01-274400.61 & Herschel973 &     & 1.404\\
    AzGS18 & J033244.02-274635.9 & J033244.01-274635.0 & J033244.01-274635.24 & Herschel403 &     & 2.688\\
    AzGS25 & J033246.83-275120.9 & J033246.84-275121.2 & J033246.83-275120.90 & Herschel961 & 178 & (1.101)\\
    AzGS9  & J033302.94-275146.9 & J033303.05-275145.8 & J033303.00-275146.27 & Herschel559 & 188 & (4.253)\\
    \hline
  \end{tabular}
\end{table*}

For an illustration of the astrophysical impact of SATMC,
we next set off to model a sample of high redshift sub-millimetre
galaxies (SMGs) using GRASIL.  SMGs are
extremely luminous in the IR, L$_{IR} \gtrsim 10^{12}\rm~L_{\odot}$
\citep{blain04,chapman05}, with a redshift distribution peaking at
z$>$2.  Given their high luminosities and IR-(sub)mm emission,
SMGs are generally believed to be very dusty systems containing
powerful starbursts \citep[SFRs$\sim
  1000\rm~M_{\odot}~yr^{-1}$][]{hughes98,coppin06,weiss09,scott10}; 
however, the exact nature of these sources is still uncertain given
the low angular resolution at (sub)mm wavelengths and relative
faintness of their optical counterparts. In \cite{johnson13}, we took
samples of AzTEC 1.1mm-detected SMGs in 3 of the most widely studied
fields, GOODS-N, GOODS-S and COSMOS, and examined their X-ray and
IR/radio counterparts for evidence of active galactic nuclei (AGNs).
Our SED analysis, performed using SATMC, showed that the IR-to-radio
SED was dominated by starburst activity whereas an AGN component,
whose IR emission was defined by the \cite{siebenmorgen04} template
library, contributed little to the overall emission but was necessary
to match the X-ray luminosity prior. The starburst templates, provided
by \cite{efstathiou00}, generally under-predicted the sub-mm flux
which suggested the presence of an extended cold dust
component. Presently, we show the results of modeling a sub-sample of
the \cite{johnson13} X-ray-detected SMGs using GRASIL, selected from
GOODS-S with additional optical and \textit{Herschel} observations.   

\subsection{Sample Selection}

For a full description of our X-ray-detected AzTEC SMG sample, we
refer the reader to \cite{johnson13}.  In summary, the fields GOODS-N,
GOODS-S and COSMOS contain the most comprehensive multi-wavelength
coverage from space and ground based facilities including XMM-Newton,
\textit{Chandra}, HST, \textit{Spitzer}, \textit{Herschel}, and VLA.
In Johnson et al., X-ray counterparts were determined for the AzTEC 
SMGs detected in each of the three fields in order to constrain the 
relative contribution from an obscured AGN.  Our preliminary analysis 
using the template libraries of \cite{siebenmorgen04} and 
\cite{efstathiou00} showed that while the X-ray detections indicated 
the presence of an AGN, the AGN component has a negligible contribution
to the mid-to-far-IR SED as the observed X-ray luminosities limit the
amount of IR emission from the AGN.  We should point out, however, that
the \cite{siebenmorgen04} AGN models assume dust reprocessing on scales
much larger than the obscuring torus which may result in colder
effective dust temperatures.  Regardless, even traditional AGN torus 
models \citep[e.g.][]{nenkova08} showed negligible IR emission when
coupled with the observed X-ray luminosities.  We may therefore assume
that our sample and the corresponding UV-through-IR SEDs are free from 
AGN contamination.  While there may be radio excess in our sample due to 
possible AGN contribution, we currently lack the information to 
accurately model the AGN, star formation, and possible radio jets.
Though we have selected a particular sub-sample of the full SMG 
population, we are confident that the results obtained with this 
sample may be applied to all SMGs as emission from star 
formation will continue to dominate in non-X-ray-detected SMGs.

For the present analysis, we limit our sample to the GOODS-S field
where we have the deepest \textit{Chandra} data to date (integrated
exposure time $\sim$4Ms) and excellent redshift coverage.  The
\textit{Chandra} data is presented in \cite{xue11} though we use the
custom source catalog of \cite{johnson13} produced through standard
\textit{Chandra} Interactive Analysis of Observations (\textsc{ciao})
reprocessing.  Though our catalog was created with a more stringent
detection criteria (false detection probability of $10^{-6}$), it
shows excellent agreement with that of Xue et al.  Utilizing 
the X-ray data provides a statistically robust counterpart selection 
due to the lower X-ray source density than comparable optical or 
near-IR catalogs; for reference, $\sim$1 source of our current sample is 
expected to result from the false overlap of X-ray and AzTEC sources.  
The X-ray data also allows for unique counterpart identification in 
the optical/IR catalogs thanks to its very small ($\sim$2\arcsec) 
positional accuracy. 

X-ray counterparts to the AzTEC GOODS-S sample \citep{scott10} of SMGs
are initially determined using a fixed search radius of 10\arcsec
which is comparable to the average search radius used in \cite{yun12}.
This results in 16 X-ray-detected AzTEC SMGs where 2 have 2 potential
X-ray counterparts.  Recently, \cite{hodge13} has provided ALMA
follow-up to 126 SMGs in the LABOCA ECDFS Submillimeter Survey
\citep[LESS,][]{weiss09} which allows an accurate counterpart
identification and examination of potential multiple sources.
However, only 4 of our 16 X-ray detected AzTEC sources are present in
the LESS catalog.  The ALMA follow-up to the 4 common sources
indicates that they are single systems (none of the potential
multiples were detected by LABOCA) whose position agrees with our
X-ray identifications to within 2\arcsec (see below).  For those AzTEC
sources with potential multiple counterparts, we treat each source
separately and make no attempt to split the AzTEC flux as we do not
have enough information to determine how the AzTEC flux may be
distributed between multiple sources. Once the X-ray counterparts have
been determined, we cross-match our X-ray catalog with publicly
available VLA \citep{kellermann08}, \textit{Herschel}
\citep{oliver12}, \textit{Spitzer} SIMPLE \citep{damen11} and
FIDEL ({\it Spitzer} PID30948, PI: M. Dickinson),
and HST \citep{giavalisco04} catalogs using a 2\arcsec search radius,
consistent with the typical positional uncertainty of the X-ray
sources; a 10\arcsec search radius was used for the \textit{Herschel}
catalog given the much larger beam-size.  

Spectroscopic and photometric redshifts were obtained by
cross-matching our X-ray catalog to that of Xue et al. which compiles
spectroscopic and photometric redshifts from 13 different groups (see
Xue et al. for more details).  For sources with photometric redshifts,
the SED fits are fixed at their photometric values.  One can include
errors on the photometric redshift as a prior (\S~3.4); however, we
only consider fixed redshifts here in order to maintain a consistent
parametrization of the SED for all sources regardless of spec-z or
photo-z quality and to avoid introducing redshift related correlations
into the fitted parameters.  To improve on the SED fitting of
\cite{johnson13} and accurately model the far-IR peak, we limit our
sample to only those with \textit{Herschel} counterparts, putting our
final sample at 12 X-ray-detected AzTEC SMGs.  A summary of these
objects is listed in Table~\ref{tab:sample}.    

\subsection{GRASIL Parametrization and SED Fitting}

\begin{table}
\caption{Description of parameters used for fitting our sample of SMGs
  with GRASIL.} 
\label{tab:grasilparam}
\begin{tabular}{lp{3cm}r}
Parameter & Description & Limiting Range \\
\hline
T$_{gal}$     & Galaxy age when 'observed' (Gyr) & [0.5,GALAGE(z)]\\
CLOUD\_RATIO & Density of molecular clouds, sets effective 1
$\mu$m optical depth (M$_{\odot}$ pc$^{-2}$)& [1.e-3,1.e7]\\
M$_{final}$  &  Total galaxy mass (gas+stars) at T$_{\rm gal}$
(M$_{\odot}$). & [1.e10,1.e13]\\
M$_{burst}$ & Total mass of stars formed during starburst
(M$_{\odot}$) & [1.e8,1.e11]\\
$\tau_{inf}$ & In-fall timescale of IGM gas onto galaxy (Gyr) & [0.1,10]\\
etastart & Escape timescale for stars from their molecular clouds
(Gyr) & [0.001,10]\\
$\nu_{sch}$ & Efficiency of Schmidt SFR law & [1.e-4,4]\\
M$_{dust}$ & Total dust mass & [1.e8,1.e10]\\
\hline
\end{tabular}
\end{table}

\begin{figure*}
\includegraphics[width=.9\textwidth]{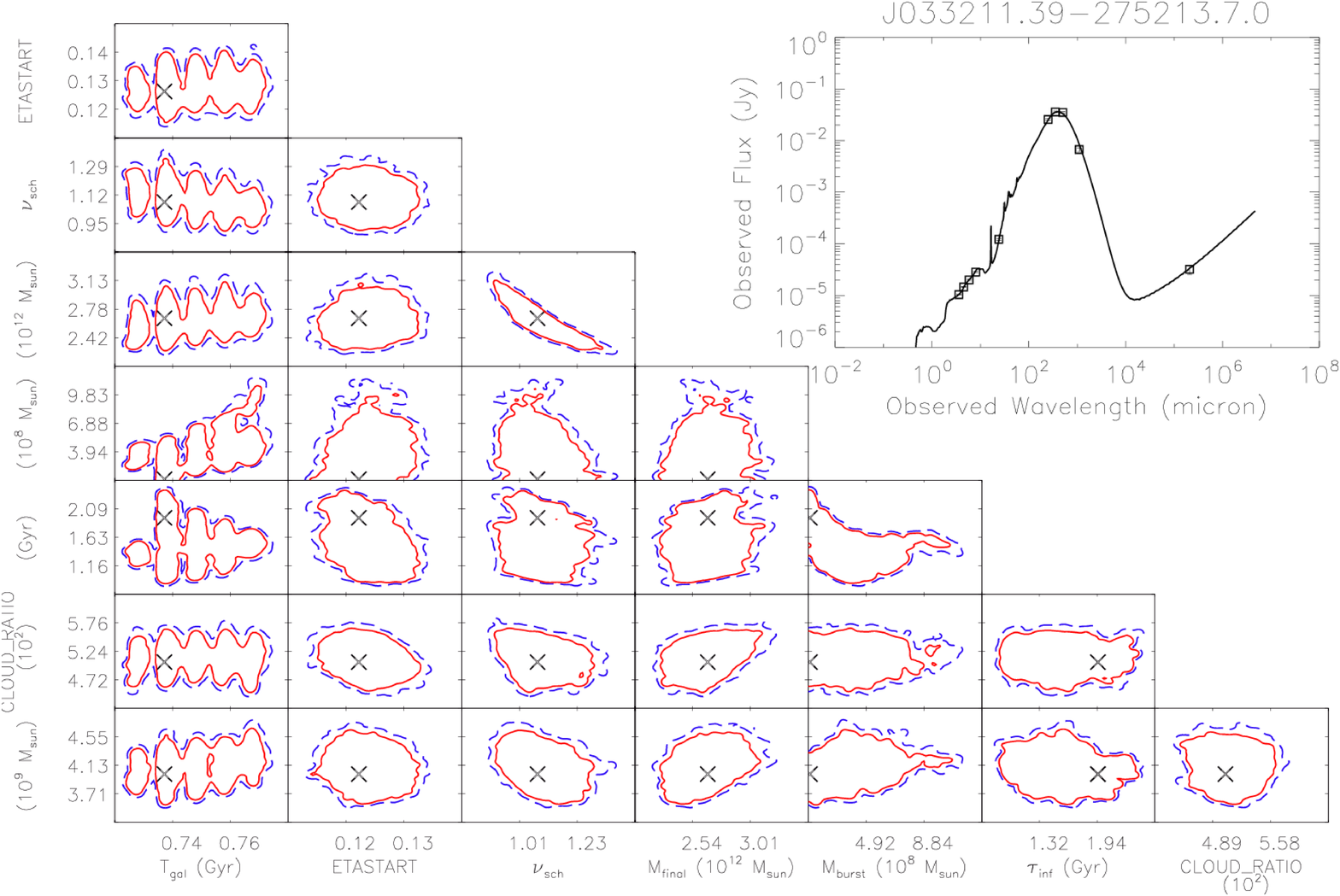}
\caption{Best fit GRASIL model to the SED of J033211.39-275213.7. 
  Shown here are the smoothed parameter-parameter likelihood 
  distributions with contours placed at the 68 percent and 90 percent
  confidence intervals.  The best fit model and observed SED are
  inlaid in the upper right corner. Though parameters in GRASIL take on a
  continuous nature, the SFRH is computed on small, discrete
  time-steps which results in the 'lumpy' confidence contours of T$_{gal}$.}
\label{fig:sample1}
\end{figure*}

GRASIL contains over 50 different parameters that control various
aspects of the output UV-to-radio SED.  These parameters cover the
dust content, geometry of dust, stars and gas, the SFRH and more.  For
our SED fitting, we choose a set of 8 free parameters, listed in
Table~\ref{tab:grasilparam}, that we find to be representative of the
major physical processes and have the most impact in our fits.  These
differ slightly from the standard input format for GRASIL but are
easier to interpret.  In addition to these free parameters, we have
set additional constraints for the SFH and dust content.  Here we
provide a brief motivation for our parametrization followed by the
application to our SMG sample.

The standard GRASIL implementation assumes that the SFH has two
components: a Schmidt law of the form $SFR(t)=\nu_{sch}{\rm
  M}_{gas}(t)^k$ and a burst which can either be constant or have an
exponential decay.  The parameters $\nu_{sch}$, $\tau_{inf}$ (rate of
new gas in-falling onto the galaxy) and M$_{final}$ (total gas mass
M$_{gas}$ and stellar mass M$_*$) control the Schmidt portion of the
SFH (we fix the exponent $k=1$ following \citealt{iglesias07}).  For
the burst component, we assume that a mass M$_{burst}$ of stars will
be created from a starburst that occurs 50 Myr prior to the final
galaxy age T$_{gal}$ with an exponential decay of 50 Myr.  Both the
burst and Schmidt components follow the same Initial Mass Function (IMF) 
which has been set to a Salpeter IMF \citep{salpeter55}.  GRASIL treats
the SFH as a 'closed-box' with no additional gas inflow outside of the 
continuous in-fall set by $\tau_{inf}$.  We therefore modify the standard 
GRASIL SFH to create a 'merger-like' SFH following the prescription 
outlined in \S~4.  In effect, the galaxy is passively evolving through 
the Schmidt law until the 'burst' occurs which deposits additional gas 
to be turned into stars.  This modified SFH aids in re-creating
major/minor mergers or sudden peaks in gas inflow/accretion where the
magnitude of the merger/in-fall event is measured with the ratio of
M$_{final}$ to M$_{burst}$, the effective 'merger ratio'.  

The second part of GRASIL is the handling of dust and radiative
transfer to produce the synthetic SED.  The primary influence to the
infrared portion of the SED is that of dust reprocessing which depends
on the dust mass M$_{dust}$ and the stellar populations.  These
stellar populations are best parametrized according to their SFH (from
above), the properties of their molecular clouds (CLOUD\_RATIO) and
the rate at which stars are able to escape their birth-sites
(etastart).  In addition to these free parameters, we fix the
dust-to-gas ratio to that of the Milky Way
\citep[$\sim$1/110][]{draine84} and assume the dust/gas/stars follow a
King radial density profile \citep[see][]{silva98}. One possible
limitation to GRASIL is that it has no prescription for any emission
from an AGN; however, as we showed in \cite{johnson13}, the AGN
contribution is likely negligible to the mid-IR-to-radio SED of our
SMG sample.  

\subsection{Do SMGs form a single population?}

\begin{table*}
\caption{Fitted parameters for our sample of 12 AzTEC SMGs using
  GRASIL.  Errors on fit parameters estimated at the 68 percent
  confidence level.}
\label{tab:grasilfits}
\begin{tabular}{lccccccccr}
Source ID & T$_{gal}$ & etastart & $\nu_{sch}$ & M$_{final}$ &
M$_{burst}$ & $\tau_{inf}$ & CLOUD\_RATIO & M$_{dust}$\\
 & Gyr & Gyr & & 10$^{11}$ M$_{\odot}$ & 10$^{9}$ M$_{\odot}$ & Gyr &
M$_{\odot}$ pc$^{-2}$ & 10$^9$ M$_{\odot}$ & ln(L)\\
\hline
J033205.34-274644.0 &  2.03$^{+ 0.09}_{- 0.35}$ &  1.07$^{+ 0.14}_{-0.35}$ &  3.81$^{+ 0.19}_{- 2.06}$ &   26.4$^{+  8.3}_{-  9.5}$ &1.91$^{+ 25.00}_{-  1.81}$ &  0.77$^{+ 1.81}_{- 0.45}$ &  1.37$^{+0.06}_{- 0.16}$e4 &   3.50$^{+  0.93}_{-  0.67}$ & -149.4\\
J033207.12-275128.6 &  1.55$^{+ 0.01}_{- 0.01}$ &  0.01$^{+ 0.03}_{-0.01}$ &  0.16$^{+ 0.05}_{- 0.15}$ &   0.46$^{+31.42}_{-  0.24}$ & 21.2$^{+  10.2}_{-  8.4}$ &  0.48$^{+ 0.11}_{- 0.27}$ & 2.08$^{+11.70}_{- 1.24}$e4 &   8.79$^{+  1.20}_{-  5.41}$  & -0.4\\
J033211.39-275213.7 &  0.74$^{+ 0.04}_{- 0.01}$ &  0.12$^{+ 0.02}_{-0.01}$ &  1.08$^{+ 0.29}_{- 0.18}$ &   26.6$^{+  6.5}_{-  4.3}$ & 0.11$^{+ 1.04}_{-  0.01}$ &  1.94$^{+ 0.46}_{- 1.05}$ &  5.04$^{+0.77}_{- 0.55}$e2 &   4.00$^{+  0.72}_{-  0.44}$  & -13.3\\
J033212.23-274620.9 &  2.62$^{+ 0.01}_{- 0.01}$ &  1.95$^{+ 0.10}_{-0.22}$ &  2.79$^{+ 0.01}_{- 0.01}$ &   5.34$^{+  0.33}_{-  0.29}$ & 0.15$^{+  1.00}_{-  0.05}$ &  0.15$^{+ 0.01}_{- 0.01}$ &  7.40$^{+0.65}_{- 0.55}$e3 &   1.71$^{+  0.46}_{-  0.25}$ &  -169.6\\
J033213.88-275600.2 &  0.82$^{+ 0.01}_{- 0.01}$ &  0.94$^{+ 9.03}_{- 0.58}$ &  0.32$^{+ 0.43}_{- 0.03}$ &   93.0$^{+  7.0}_{-  48.4}$ &   9.66$^{+  3.36}_{-  9.11}$ &  8.12$^{+ 1.87}_{- 5.83}$ &  1.52$^{+ 0.20}_{- 0.10}$e2 &   1.11$^{+  0.32}_{-  0.27}$  & -80.3\\
J033215.32-275037.6 &  1.16$^{+ 2.03}_{- 5.00}$ &  7.48$^{+ 2.52}_{- 6.62}$ &  0.02$^{+ 0.01}_{- 0.01}$ &   16.8$^{+  0.8}_{-  0.9}$ &   0.10$^{+  0.01}_{-  0.01}$ &  0.15$^{+ 0.01}_{- 0.01}$ & 13.76$^{+ 1.01}_{- 1.33}$ &   3.97$^{+  0.41}_{-  0.30}$  & -1816.4\\
J033222.17-274811.6 &  2.79$^{+ 0.01}_{- 0.01}$ &  0.21$^{+ 0.05}_{- 0.01}$ &  1.63$^{+ 0.01}_{- 0.08}$ &   0.201$^{+  0.02}_{-  0.01}$ &   29.5$^{+  2.9}_{-  1.4}$ &  1.83$^{+ 0.01}_{- 0.05}$ &  4.68$^{+ 0.18}_{- 0.13}$e2 &   5.75$^{+  0.72}_{-  0.05}$ & -155.8\\
J033222.56-274815.0 &  1.41$^{+ 0.01}_{- 0.04}$ &  0.50$^{+ 0.44}_{- 0.18}$ &  0.15$^{+ 0.18}_{- 0.05}$ &   24.0$^{+  10.2}_{-  12.6}$ &   4.60$^{+  5.75}_{-  1.90}$ &  2.19$^{+ 3.10}_{- 0.51}$ &  1.71$^{+ 0.35}_{- 0.25}$e2 &   0.87$^{+  0.25}_{-  0.27}$ & -16.9\\
J033238.01-274401.2 &  2.02$^{+ 0.01}_{- 0.01}$ &  3.83$^{+ 6.17}_{- 2.16}$ &  3.48$^{+ 0.06}_{- 0.65}$ &   3.52$^{+  0.18}_{-  0.40}$ &   0.11$^{+  0.59}_{-  0.01}$ &  0.28$^{+ 0.03}_{- 0.03}$ &  1.69$^{+ 0.13}_{- 0.16}$e3 &   9.76$^{+  0.24}_{-  2.17}$ & -31.6\\
J033244.02-274635.9 &  1.03$^{+ 0.01}_{- 0.01}$ &  0.28$^{+ 0.05}_{- 0.02}$ &  0.18$^{+ 0.01}_{- 0.07}$ &   29.8$^{+  16.2}_{-  1.7}$ &   15.8$^{+  1.6}_{-  5.3}$ &  0.22$^{+ 0.01}_{- 0.10}$ &  7.18$^{+ 0.49}_{- 0.53}$e2 &   6.02$^{+  0.81}_{-  0.94}$  & -69.3\\
J033246.83-275120.9 &  4.01$^{+ 2.05}_{- 5.00}$ &  2.78$^{+ 0.27}_{- 0.12}$ &  0.54$^{+ 0.01}_{- 0.01}$ &   2.95$^{+  0.07}_{-  0.07}$ &   0.84$^{+  0.02}_{-  0.02}$ &  0.75$^{+ 0.01}_{- 0.01}$ &  9.95$^{+ 0.20}_{- 0.39}$e2 &   3.16$^{+  0.00}_{-  0.09}$ & -772.4\\
J033302.94-275146.9 &  1.00$^{+ 5.28}_{- 5.00}$ &  1.27$^{+ 8.70}_{- 0.65}$ &  0.46$^{+ 0.11}_{- 0.03}$ &   26.3$^{+  1.3}_{-  3.7}$ &   2.24$^{+  1.01}_{-  1.11}$ &  0.19$^{+ 0.15}_{- 0.05}$ &  1.96$^{+ 0.11}_{- 0.15}$e2 &   1.32$^{+  0.28}_{-  0.35}$ & -283.3\\
\hline
\end{tabular}
\end{table*}

\begin{table*}
\caption{Derived attributes from the fitted parameters of
  Table~\ref{tab:grasilfits}. SFR and stellar mass M$_*$ are provided
  as outputs from the GRASIL SFH calculation. The SFRs reported are
  the SFRs averaged over the age of the burst (50 Myr prior to
  T$_{gal}$) with stellar masses given at T$_{gal}$.  For
  comparison, we also include the SFR as derived from the FIR
  luminosity (SFR$_{FIR}$) of the best fit model following
  Kennicutt 1998.  Specific star formation rates (SSFRs) are simply
  SFR/M$_*$.  Errors on SFR and M$_*$ are given at the 68 percent
  confidence level.}  
\label{tab:grasilder}
\begin{tabular}{lrrrr}
Source ID & SFR & SFR$_{\rm FIR}$ & M$_*$ & SSFR \\
 & M$_{\odot}$ yr$^{-1}$ & M$_{\odot}$ yr$^{-1}$ & 10$^{11}$ M$_{\odot}$ & Gyr$^{-1}$ \\
\hline
J033205.34-274644.0 &  950.8$^{+419.3}_{-227.6}$ & 1230.0 & 23.14$^{+8.39}_{-9.65}$ &  0.41$^{+0.42}_{-0.18}$ \\
J033207.12-275128.6 &  428.8$^{+235.5}_{-140.7}$ &  182.5 & 0.23$^{+0.38}_{-0.15}$ & 19.26$^{+42.82}_{-13.20}$ \\
J033211.39-275213.7 & 2086.5$^{+202.2}_{- 90.6}$ & 1892.8 & 6.80$^{+0.52}_{-0.33}$ &  3.07$^{+0.19}_{-0.13}$ \\
J033212.23-274620.9 &   53.2$^{+ 19.6}_{-  2.6}$ &   84.4 & 4.86$^{+0.29}_{-0.27}$ &  0.11$^{+0.04}_{-0.01}$ \\
J033213.88-275600.2 & 2829.0$^{+166.7}_{-304.7}$ & 1966.0 & 8.50$^{+0.97}_{-0.45}$ &  3.33$^{+0.10}_{-0.42}$ \\
J033215.32-275037.6 &   30.3$^{+  0.6}_{-  0.5}$ &   18.1 & 0.21$^{+0.01}_{-0.01}$ &  1.46$^{+0.01}_{-0.01}$ \\
J033222.17-274811.6 &  518.0$^{+217.8}_{-  0.4}$ &  187.1 & 0.36$^{+0.02}_{-0.04}$ & 14.60$^{+6.59}_{-0.40}$ \\
J033222.56-274815.0 &  379.2$^{+139.2}_{- 13.9}$ &  308.2 & 1.86$^{+0.11}_{-0.28}$ &  2.04$^{+1.05}_{-0.08}$ \\
J033238.01-274401.2 &   59.6$^{+ 10.5}_{-  4.5}$ &   64.0 & 3.21$^{+0.17}_{-0.38}$ &  0.19$^{+0.04}_{-0.01}$ \\
J033244.02-274635.9 &  745.7$^{+ 76.7}_{- 75.9}$ &  458.4 & 3.03$^{+0.40}_{-0.13}$ &  2.46$^{+0.24}_{-0.36}$ \\
J033246.83-275120.9 &   74.9$^{+  0.2}_{-  5.2}$ &   50.0 & 1.83$^{+0.05}_{-0.04}$ &  0.41$^{+0.01}_{-0.03}$ \\
J033302.94-275146.9 &  955.5$^{+ 24.6}_{- 18.1}$ & 1018.3 & 6.09$^{+0.11}_{-0.54}$ &  1.57$^{+0.19}_{-0.06}$ \\
\hline
\end{tabular}
\end{table*}

\begin{figure*}
\includegraphics[width=0.9\textwidth]{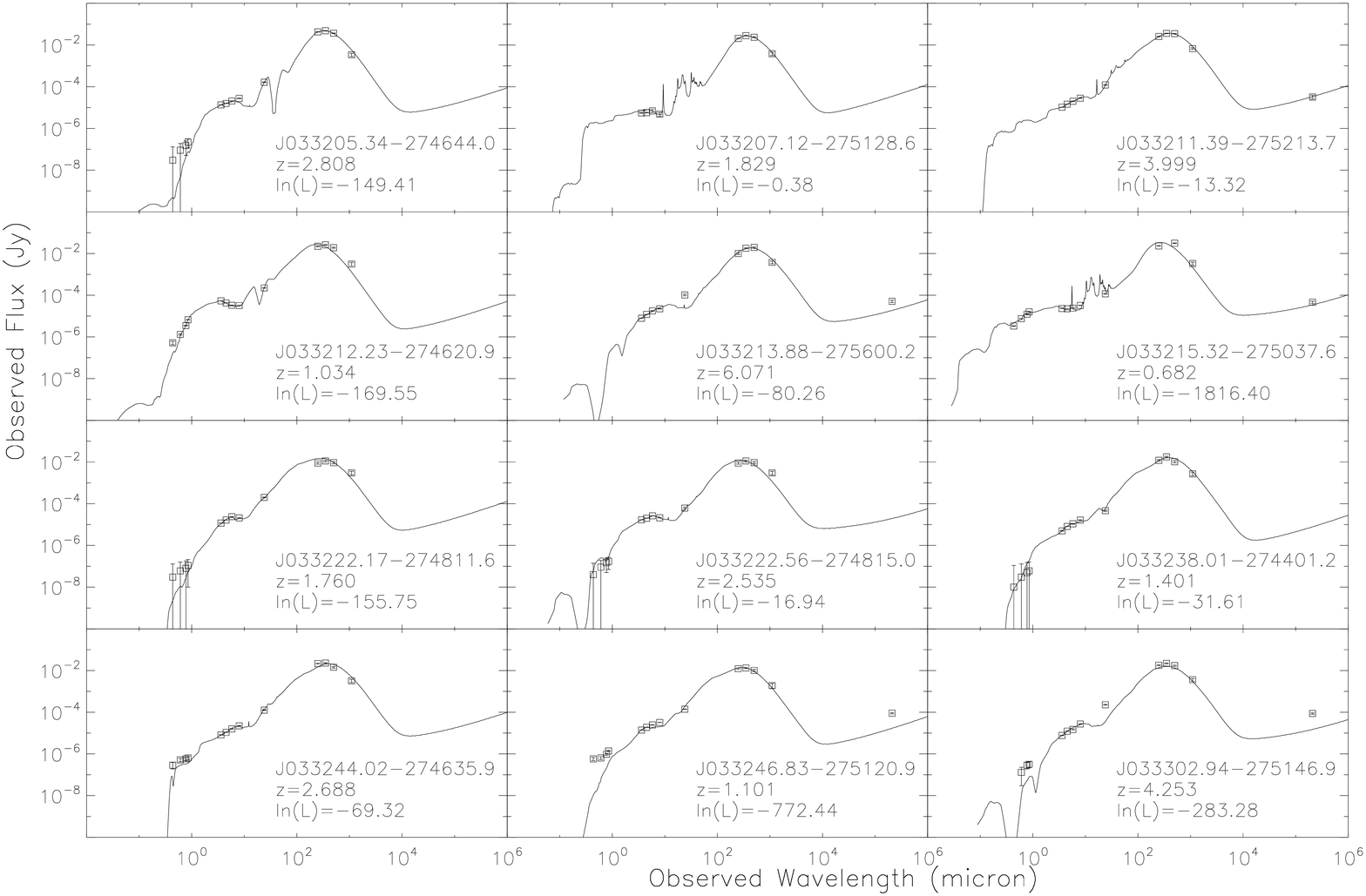}
\caption{Best fit GRASIL models to the SMG sample.  Included in each
  panel are the source ID, redshift used during the fitting (see
  Table~\ref{tab:sample}), and the ln(L) value of the fit.}
\label{fig:grasilseds}
\end{figure*}

With our GRASIL parameter space as defined above, it is
straight-forward to fit our SMGs sample with SATMC.  A sample of 
the final SED produced by SATMC is shown in Figure~\ref{fig:sample1} 
with a summary of the fitting results in Figure~\ref{fig:grasilseds} 
and Tables~\ref{tab:grasilfits} and \ref{tab:grasilder}. An important 
aspect to note from the parameter-parameter covariances is that 
many parameters show tight constraints and large
degeneracies generally do not exist in our adopted parametrization.
The strongest correlations exist between M$_{final}$ and $\nu_{sch}$
which together set the stellar mass M$_*$.  In the observed SEDs, the
stellar mass is measured through the stellar bump which is most
prominent in the IRAC bands.  As the IRAC observations have very small
flux errors ($\lesssim$1\%), this limits parameter space to the narrow
range of  M$_{final}$ and $\nu_{sch}$ capable of reproducing the
observed fluxes.    

\begin{figure*}
\includegraphics[width=0.45\textwidth]{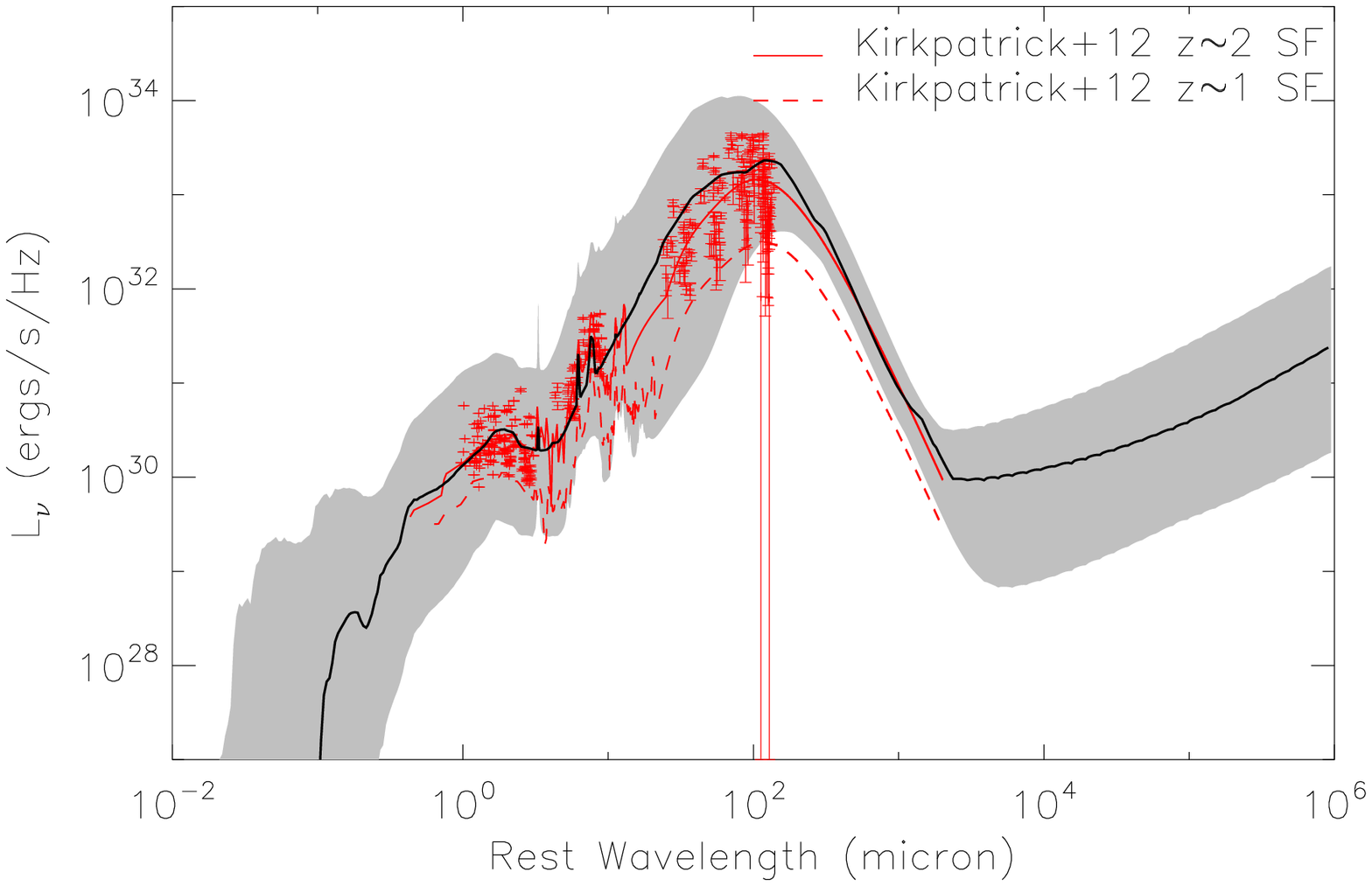}
\includegraphics[width=0.45\textwidth]{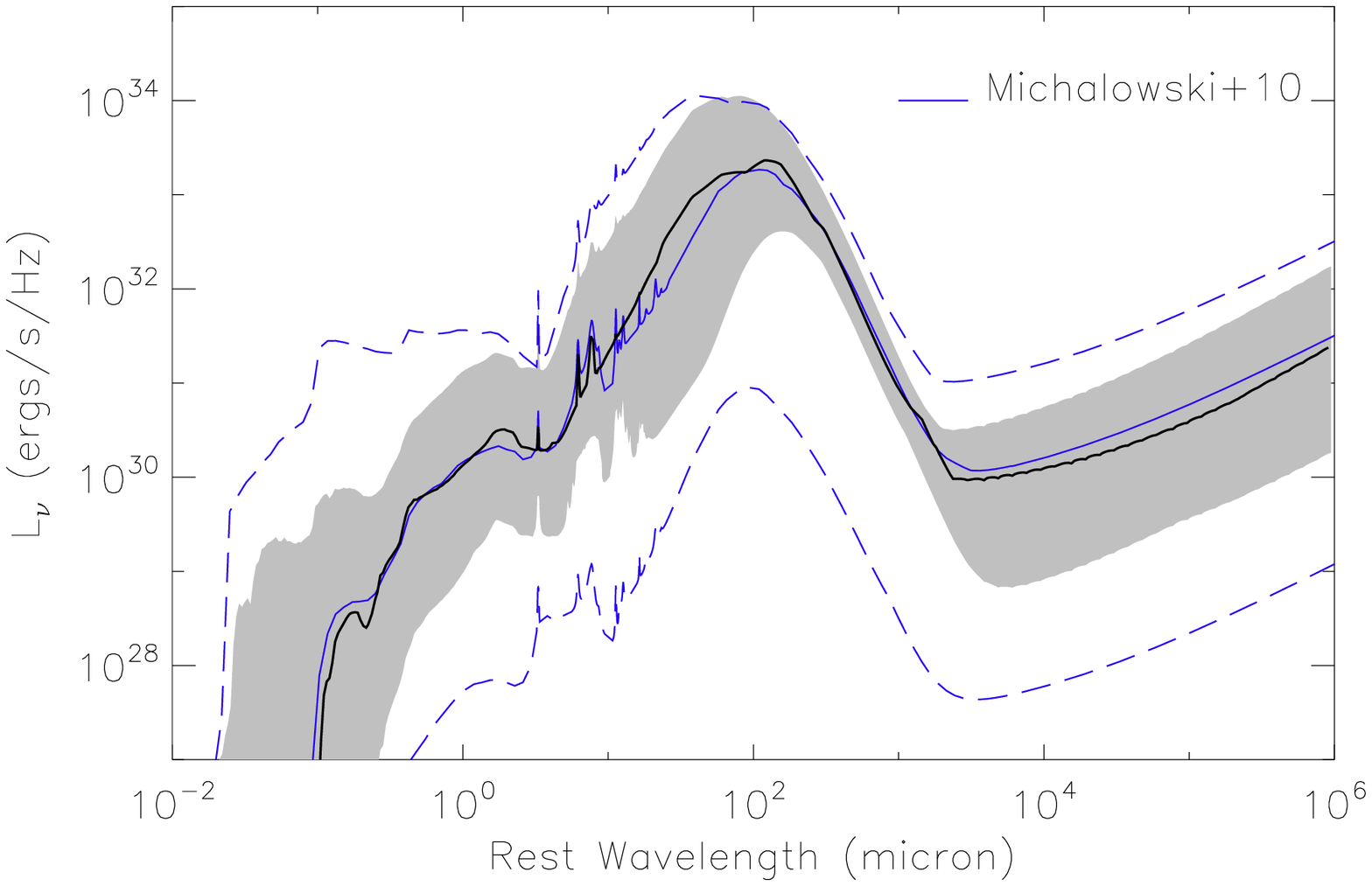}
\caption{Compiled SEDs for the SMG sample.  The median SED is given by
  the solid black line with the grey shaded region indicating the
  dynamic range between individual SEDs. {\it Left:} The composite
  SEDs overlaid with the average star-formation templates of
  \citealt{kirkpatrick12}.  The photometry used in constructing the
  $z\sim2$ SF sample has been included to demonstrate the scatter in the
  data.  {\it Right:}  The composite SEDs overlaid with median SMG
  template of \citealt{michalowski10} with the dashed lines
  encompassing the dispersion in the Mickalowski et al. models.  Note
  that these models have \textit{not} been re-normalized to fit our
  model set.} 
\label{fig:compseds}
\end{figure*}

When examining a population of sources, it is often helpful to 
view the SEDs as a composite and construct an empirical median template.
This allows one to easily see trends common in the population including
the approximate flux ranges covered by the models and dominate aspects
of the SEDs.  Figure~\ref{fig:compseds} shows our composite rest-frame
GRASIL models along with the median SED of \cite{michalowski10}, who
fit the SEDs of SMGs using a grid developed from GRASIL, and the
composite star-forming SEDs of \cite{kirkpatrick12}, derived from
$z\sim1$ and $z\sim2$ sources in GOODS-N and ECDFS with
\textit{Spitzer} IRS spectroscopy.  Though Michalowski et al. and
Kirkpatrick et al. normalize their fits at specific wavelengths to
construct the composite, we have applied no re-normalization to match
our SED set.  The agreement between our
SEDs with those of Mickalowski et al. and Kirkpatrick et al. is
immediately apparent. From Kirkpatrick et al., our composite SED best
agrees with the $z\sim2$ star-forming galaxy; not surprising as SMGs
peak around $z\sim2.5$.  Our composite also shows excellent agreement
with the median Michalowski et al. SED. There are some minor
discrepancies between the composites though they are likely attributed
to differences in parametrizing the SED.  Compared to the Kirkpatrick
et al. composites, our SFRs and stellar mass estimates are higher by a
factor of $\sim$2 which can readily be explained by the SFH adopted.
Regardless, our composite SED is not only able to recover the medians
of Kirkpatrick et al. and Michalowski et al. but also encompass the
scatter in the observations and dispersion in similar template sets.

\begin{figure*}
\includegraphics[width=.9\textwidth]{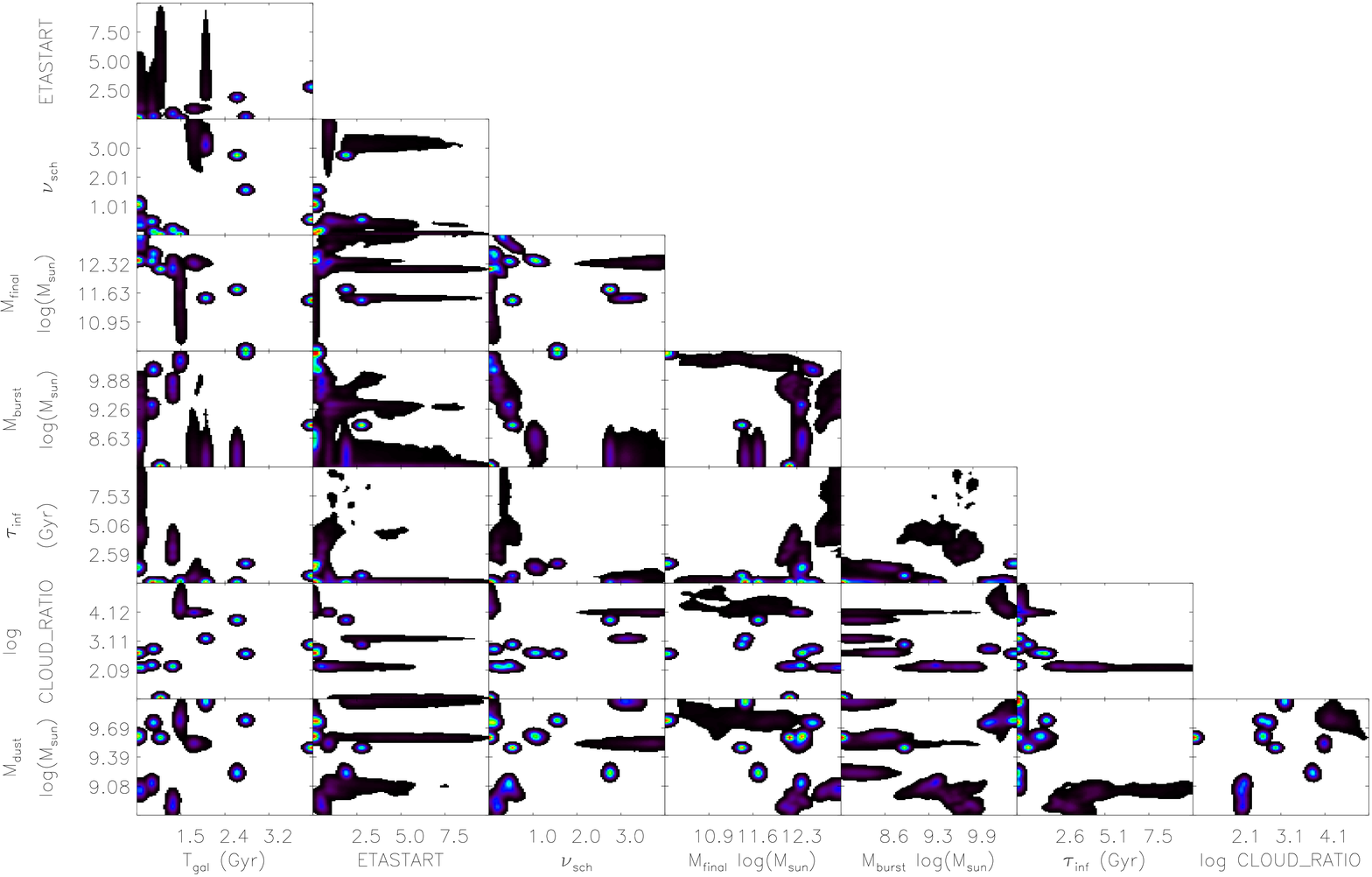}
\caption{Compiled histogram distribution of accepted steps from all
  fits.  Note that the histogram distributions differ slightly from
  the likelihood distributions but maintain the same general
  characteristics due to the multi-dimensional nature of the data
  volume.} 
\label{fig:compfit}
\end{figure*}

Despite the similar shapes and well defined empirical median of our
fitted SEDs, the SMGs in our sample are not heterogeneous in nature and,
in fact, occupy large regions of parameter space.  Fig.~\ref{fig:compfit}
shows that while some areas of parameter space are well constrained
(e.g. $\tau_{inf}$ versus T$_{gal}$), there is no single set of 
parameters that may describe our sample as an individual population.
Of all our adopted parameters, M$_{dust}$ shows the tightest constraints
(Table~\ref{tab:grasilparam}). This is primarily due to the inclusion
of the (sub-)mm observations which are highly sensitive to the total
dust mass. Though the parameters $\nu_{sch}$ and M$_{total}$
had shown tight correlations in individual fits, the composite fits
show no signs of the original correlations.  

\subsection{SMGs and the 'Main-Sequence' of Galaxy Formation}

It is often believed that SMGs must result from powerful starbursts 
possibility triggered by major mergers.  However, these results
show a wide range in derived SFRs ($\sim$20-2000 M$_{\odot}$ yr$^{-1}$)
with many below the $\sim$1000 M$_{\odot}$ yr$^{-1}$ typically assumed for 
SMGs.  Furthermore, the 'merger ratio' (M$_{burst}$/M$_{final}$) would 
also suggest that major mergers are uncommon.  This is in agreement with
\cite{hayward11} who find that starbursts are highly inefficient at
boosting the sub-mm flux (e.g. a $\gtrsim$16$\times$ boost in SFR
produces $\lesssim$2$\times$ increased sub-mm flux).  Indeed, the
common trend in our sample is that the bulk of the stellar mass is
built from the Schmidt component rather than a 'merger' triggering a
starburst.  A more detailed study including a larger sample is needed
to further clarify this point.  The sources with low SFR could 
result from source blending giving rise to an enhanced sub-mm flux 
as also suggested by Hayward et al.  \cite{karim13} have shown that
the brightest LABOCA sources ($S_{850\mu m}\gtrsim 12$ mJy) are composed
of multiple, fainter SMGs.  For our AzTEC SMG sample, however, the
corresponding bright flux limit is $\sim$6.7 mJy \citep[assuming 
  R=$S_{850\mu m}/S_{1.1mm}=1.8$;][]{chapman09} which only applies to the 
source J033211.39-275213.7, a source with no multiple X-ray detection
and derived SFR$\sim$2000 M$_{\odot}$ yr$^{-1}$.  Furthermore, only 2 to 
3 out of 12 of our SMGs have multiple potential counterparts 
(Table~\ref{tab:sample}), none of which correspond to a low SFR source.  
While source blending is certainly problematic for correctly interpreting 
multi-wavelength SED fits, it does not have a significant influence in 
our sample.

\begin{figure}
\includegraphics[width=0.45\textwidth]{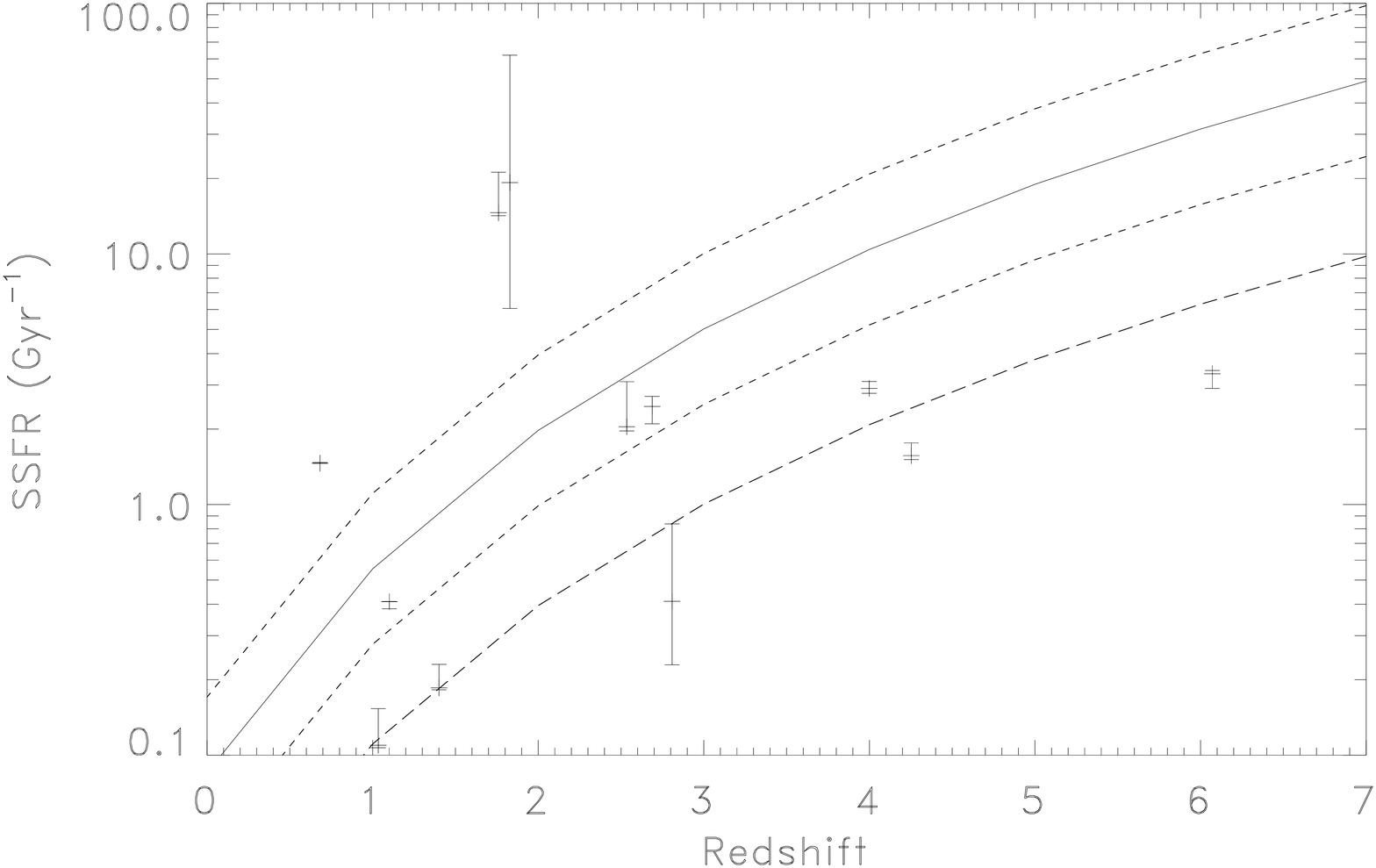}
\caption{SSFR as a function of
  redshift.  The area within the dashed lines represents the
  main-sequence of galaxies from \citealt{elbaz11} with galaxies
  above this limit defined as starbursts.  The lower long-dashed line
  represents the approximate main-sequence relation for our sample
  accounting for the factor of $\sim$5 difference in stellar masses
  between the Elbaz et al. sample and those presented here.}
\label{fig:sfrmstar}
\end{figure}

\begin{table*}
\caption{Comparison of fitted parameters for the source
  J033213.88-275600.2 using a Salpeter and Chabrier IMF.}
\label{tab:imffit}
\begin{tabular}{lccccccccr}
IMF & T$_{gal}$ & etastart & $\nu_{sch}$ & M$_{final}$ &
M$_{burst}$ & $\tau_{inf}$ & CLOUD\_RATIO & M$_{dust}$ & ln(L)\\
 & Gyr & Gyr & & 10$^{11}$ M$_{\odot}$ & 10$^{9}$ M$_{\odot}$ & Gyr &
M$_{\odot}$ pc$^{-2}$ & 10$^9$ M$_{\odot}$\\
\hline
Salpeter &  0.82$^{+ 0.01}_{- 0.01}$ &  0.94$^{+ 9.03}_{- 0.58}$ &  0.32$^{+ 0.43}_{- 0.03}$ &   93.0$^{+  7.0}_{-  48.4}$ &   9.66$^{+  3.36}_{-  9.11}$ &  8.12$^{+ 1.87}_{- 5.83}$ &  1.52$^{+ 0.20}_{- 0.10}$e2 &   1.11$^{+  0.32}_{-  0.27}$ & -80.3\\
Chabrier &  0.90$^{+ 0.01}_{- 0.01}$ &  0.94$^{+ 9.01}_{- 0.38}$ &  0.52$^{+ 0.06}_{- 0.11}$ &   88.3$^{+  11.7}_{-  7.2}$ &   1.16$^{+23.69}_{-  1.06}$ &  0.19$^{+ 0.15}_{- 0.09}$ &  1.30$^{+ 0.17}_{-0.03}$e2 &   1.44$^{+  0.20}_{-  0.52}$  & -102.9\\
\hline
\end{tabular}
\end{table*}

Expanding on this point, we show the specific star formation 
rate (SSFR=SFR/M$_*$) with respect to redshift for our SMG sample 
in Fig.~\ref{fig:sfrmstar}.  \cite{elbaz11} have suggested an 
evolution in SSFR for quiescent and starbursting galaxies which follow 
the general forms SSFR=26$\times t^{-2.2}$ Gyr$^{-1}$ and 
SSFR$>56\times t^{-2.2}$ Gyr$^{-1}$, where $t$ is the cosmic time, for
quiescent or 'main-sequence' galaxies and starbursts, respectively. 
We should point out, however, that the stellar masses used to derive
the Elbaz et al. relation are on order $\sim 5\times$ lower than
stellar masses in our sample.  As a result, only 3 of our SMGs lie
above the 'main-sequence' relation of Elbaz et al.  Despite this
discrepancy, our SFRs and stellar masses agree with
\cite{michalowski10} and the predictions of  \cite{hayward11};
differences in stellar mass derivations can be attributed to
differences in the parametrization of the SFRH, IMF (see \S~5.4) and
stellar evolution tracks \citep[see also][]{michalowski12}. Given the
uncertainties in determining the stellar mass, we caution against
over-interpretation of the SSFRs.  However, adjusting for the
relative differences in stellar masses, we see that the majority
of our sample lie above or near the main-sequence relation while
spanning a large range of SSFRs.  Overall, these results suggest that
there is not a single population that defines our SMG sample; instead,
they are homogeneous in origin composed of a mixture of starburst and
quiescent galaxies of various ages and SFRHs.

\subsection{Comparing Model Parametrizations}

\begin{table}
\caption{Comparison of derived parameters for J033213.88-275600.2
  using Salpeter and Chabrier IMFs.}
\label{tab:imfder}
\begin{tabular}{lrrr}
IMF & SFR & M$_{*}$ & SSFR \\
 & M$_{\odot}$ yr$^{-1}$ & 10$^{11}$ M$_{\odot}$ & Gyr$^{-1}$ \\
\hline
Salpeter & 2829.0$^{+166.7}_{-304.7}$ &  8.50$^{+0.97}_{-0.45}$ &   3.33 \\
Chabrier & 3893.4$^{+192.1}_{-122.5}$ & 13.47$^{+1.79}_{-1.94}$ &   2.89 \\
\hline
\end{tabular}
\end{table}

\begin{figure}
\includegraphics[width=0.45\textwidth]{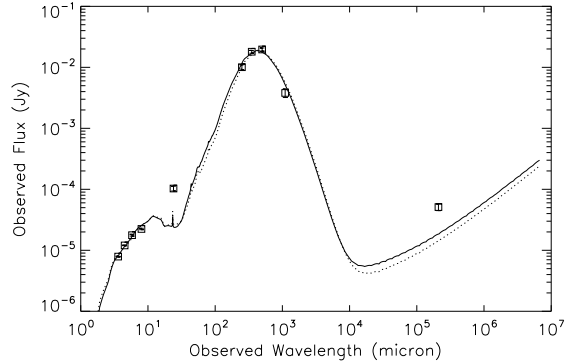}
\caption{Best fit GRASIL SEDs for J033213.88-275600.2 using Salpeter
  (solid) and Chabrier (dashed) IMFs.}
\label{fig:imfcompare}
\end{figure}

When developing SED models and deriving the associated parameters,
there are certain attributes that must remain fixed in order to keep
a consistent parametrization.  Examples of these attributes include the 
IMF, spatial properties like the disk/bulge scale heights and radial 
distribution, and the form of the SFH.  Altering any of these properties 
changes the very nature of the models and is in no way continuous for
MCMC samplers.  In Bayesian statistics, it is possible
to compare models of different parametrizations through the {\it Bayes
factor} to determine which model offers the 'best' fit
\citep[e.g.][]{jefferys61,weinberg12,weinberg13}.  The Bayes factor
follows from Bayes' Theorem and is derived from the ratio of the
posteriors for the two models $M_1$ and $M_2$:
\begin{equation}
\frac{P(M_1|{\bf D})}{P(M_2|{\bf
    D})}=\frac{P(M_1)}{P(M_2)}\frac{P({\bf D}|M_1)}{P({\bf D}|M_2)}.
\end{equation}
The Bayes factor is defined as the second term on the right-hand side;
the first term is the ratio of the prior probability of each model and
is typically set to unity as the models are considered under equal
weight.  Explicit calculation of the Bayes factor ($B_{12}$) involves
integrating the likelihood distributions of each model ($L_1({\bf
  D}|\theta_1)$ and $L_2({\bf D}|\theta_2)$) with the prior parameter
distributions ($\pi_1(\theta_1)$ and $\pi_2(\theta_2)$) according to
\begin{equation}
B_{12}=\frac{P({\bf D}|M_1)}{P({\bf D}|M_2)}\frac{\int \pi_1(\theta_1|M_1)
L_1({\bf D}|\theta_1,M_1){\rm d}\theta_1}{\int \pi_2(\theta_2|M_2)
L_2({\bf D}|\theta_2,M_2){\rm d}\theta_2)}.
\end{equation}
While there are MCMC and Bayesian based tools available
that compute the Bayes factor \citep[e.g. the Bayesian
  Inference Engine BIE,][]{weinberg12}, such methods are not currently
available in SATMC.  Instead, we may follow a similar prescription
by repeating the SED fits after changing the desired parametrization(s).
We provide a brief example of this process by comparing the results
obtained with the Salpeter IMF used in our initial fits and those 
obtained with a Chabrier IMF \citep{chabrier03}.  Tables~\ref{tab:imffit}
\& \ref{tab:imfder} and show that for the source J033213.88-275600.2 
the fitted parameters remain nearly constant except for $\tau_{inf}$ 
which causes an increase in the derived SFRs and stellar mass while
the fitted SEDs are nearly identical (Figure~\ref{fig:imfcompare}).
We caution, however, that without results from a full sample and direct 
computation of the Bayes factor, we can not differentiate between the two 
IMFs at this time.  Without the Bayes factors, one may include 
secondary observations, i.e. spectroscopy and SED independent estimates of 
fitted and/or derived parameters, to derive priors and determine which, if 
any, of the fits best describe all observations.  

\subsection{Application of Additional Observations to SED Fits}

\begin{figure*}
\includegraphics[width=0.9\textwidth]{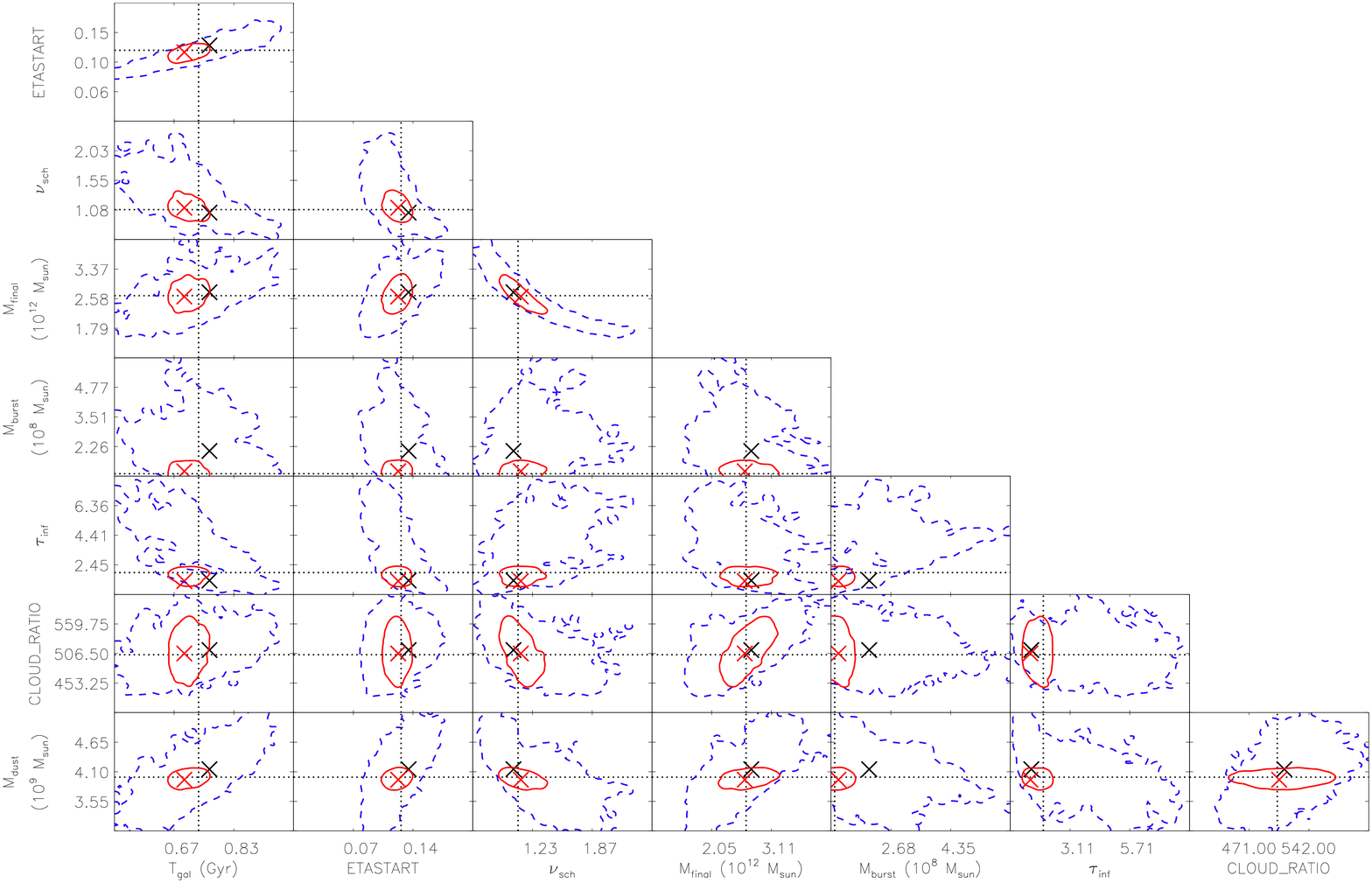}
\caption{68 percent error contours for a simulated GRASIL SED based on
  the fitted spectrum of J033211.39-275213.7 at z$\sim 4$.  The 
  contours are shown for the original AzTEC photometric errors (blue dashed) 
  and improved by a factor of 10 (red solid).  The location of the maximum 
  likelihoods for each case are marked by the X's with values used in 
  construction of the simulated SEDs marked by the dotted lines.  For 
  the high-z simulation, improving the AzTEC signal-to-noise provides 
  significantly greater constraints on parameter space.} 
\label{fig:errorsim}
\end{figure*}

As mentioned in \S~2.2, whenever an SED synthesis routine is used in
SATMC, the SEDs for each step are saved along with its location in
parameter space and likelihood.  Using these 'SED steps', one can
examine the influence of additional or improved photometric data
simply by referring back to Eqns. 5 \& 8 without having to re-run the
entire MCMC fit.  This can be exceptionally useful given the long
typical run time of a single fit and is best used for examining the
influence on parameter space when including low signal-to-noise
observations (e.g. LABOCA) or observations one believes will truncate
regions of parameter space (e.g. priors on parameters).  To examine the
full impact of improved or additional observations, it is necessary to
repeat the fits as the sampling of likelihood space is wholly
dependent on the initial set of observations and relative uncertainties.  
As an example, we have created simulated GRASIL SEDs based on the 
best fit parameters of J033211.39-275213.7 ($z\sim4$).  The simulated SED
is produced using the same formalism and parametrization described in 
\S~4 \& 5.2 and has been sampled at wavelengths corresponding to the
actual observations (3.6$\mu$m, 4.5$\mu$m, 5.8$\mu$m, 8.0$\mu$m, 
24$\mu$m, 250$\mu$m, 350$\mu$m, 500$\mu$m, 1.1mm). Each band is then 
assigned the measured flux uncertainties.  For one simulation, we have 
increased the AzTEC signal-to-noise by a factor of 10 
($\sigma_{1.1mm}\sim0.1$ mJy) -- similar to the level of improvement 
expected with LMT, ALMA and future sub-mm telescopes -- whereas the 
control simulation maintains the original AzTEC signal-to-noise 
($\sigma_{1.1mm}\sim$1 mJy). Figure~\ref{fig:errorsim} shows the 68 
percent contours for the GRASIL fits to the simulated SEDs where we 
immediately see tighter constraints on fitted parameters for the 
improved AzTEC photometry.  Referring back to 
Figure~\ref{fig:grasilseds}, we see that the AzTEC photometry is one 
of the few points beyond the thermal dust peak and along the 
Rayleigh-Jeans tail of the dust blackbody distribution.  At lower 
redshifts, the {\it Herschel}/SPIRE bands become more prominent in
the beyond the dusk peak which serves to diminish the impact of
improved (sub-)mm photometry.  For the most distant galaxies, however,
high signal-to-noise (sub-)mm photometry is instrumental for fully 
constraining the SED shape and its driving parameters, namely its SFRH 
($T_{gal}$, $M_{total}$, $M_{burst}$, $\nu_{sch}$, $\tau_{inf}$) and dust 
content $M_{dust}$.  The future observations of SMGs with LMT 
and ALMA will therefore be paramount to our understanding of high-redshift 
galaxies. 

\section{Summary}

We have presented the general purpose MCMC-based SED fitting tool
SATMC.  Utilizing MCMC algorithms, the code is able to take any set of
SED templates or models of the user's choice and return the requested
best fit parameter estimates in addition to the posterior parameter
distribution which can be used to construct confidence levels and
unveil parameter correlations.  In testament to SATMC's flexibility,
we have provide a series of test cases comparing SATMC with
traditional least-squares methods where SATMC recovers best fit values
similar to those from traditional methods.  Furthermore, we show
that SATMC is capable of reproducing the set of input parameters from
a simulated SED which serves to prove SATMC's adaptability and
reliability.   

Highlighting SATMC's scientific value, we also provide the best-fit
GRASIL models to a sample of AzTEC submillimetre galaxies in the
GOODS-S field.  These fits indicate that while some parameters exhibit
obvious and strong correlations in individual SED fits, such
correlations are absent when constructing the population.
Furthermore, these sources appear to span a wide dynamic range of
fitted and derived parameters, specifically when considering their
derived SFRs ($\sim$30-3000 M$_{\odot}$ yr$^{-1}$) and stellar masses
($\sim 10^{10}-10^{12}$ M$_{\odot}$); these results agree both with
the previous GRASIL-based SEDs of \cite{michalowski10} and simulations
of SMGs \citep{hayward11}.  We also provide a brief demonstration of
the impact of improved photometric data on the best fit results.  
Increasing the signal-to-noise of the AzTEC photometry provides
significantly greater constraints on fitted parameters and highlights
the need to obtain deeper (sub)-mm data.  As observations, the
complexity of models and the desire to obtain more detailed
information from model fits increase, the approach of tools like SATMC
will be required in both observational and theoretical astronomy in
order to refine our understanding of the Universe and its
constituents.  

\section*{Acknowledgments}

We would like to thank V. Acquaviva, the CANDELS collaboration and
A. Kirkpatrick for their assistance in developing and providing
data sets for testing and improving our MCMC algorithm.  We
would also like to thank the anonymous reviewer for their helpful
comments on improving the manuscript.  This work has
been funded in part by the North East Alliance under the National
Science Foundation (NSF) grant HRD 0450339, NSF grants AST-0907952 and
AST-0838222 and CXO grant SAO SP1-12003X. K.S. Scott is supported by
the National Radio Astronomy Observatory, which is a facility of the
NSF operated under cooperative agreement by Associated Universities,
Inc.

\end{document}